\documentclass[a4paper,11pt]{article}
\pdfoutput=1
\usepackage{jcappub}
\usepackage[T1]{fontenc}
\usepackage{amsmath,amssymb,mathtools,bm}
\usepackage{booktabs,array}
\usepackage{graphicx}
\usepackage{enumitem}
\numberwithin{equation}{section}

\newcommand{\mpl}{m_{\rm pl}}

\newcommand{\rhoMH}{\rho_M^H}

\newcommand{\GaM}{\Gamma_M}
\newcommand{\GaC}{\Gamma_C}
\newcommand{\Gde}{\Gamma_M^{\rm de}}

\newcommand{\meff}{m_{\rm eff}}

\newcommand{\JJ}{\mathbf{J}}
\newcommand{\kk}{\mathbf{k}}
\newcommand{\xx}{\mathbf{x}}
\newcommand{\dd}{\mathrm{d}}
\newcommand{\Ocal}{\mathcal{O}}

\title{\boldmath Particle--antiparticle asymmetry from charge separation at expanding horizons}

\author{She-Sheng Xue}
\affiliation{ICRANet, Piazzale della Repubblica 10, 65122 Pescara, Italy
\\ Physics Department, Sapienza University of Rome,
Rome, Italy\\INFN, Sezione di Perugia,
Perugia, Italy
\\ICTP-AP, University of Chinese Academy of Sciences, Beijing, China
}
\emailAdd{shesheng.xue@icranet.org}

\abstract{
In a Friedmann universe with a time-varying cosmological term, massive particle--antiparticle pairs are produced continuously from the spacetime vacuum, and their members are
separated by horizons. Along the cosmic history that explains $n_s$, $A_s$, $r$ and the low-$\ell$ power anomaly, we show how this separation generates the observed asymmetry. Horizons absorb
the charge they separate, since gravity does not conserve global charges. Charge separation at the many horizons inside today's Hubble volume then acts as a volume violation of the
U(1) charge, at a rate $\Gamma_C=\kappa_C\Gamma_M$ proportional to pair production. The sign requires a C- and CP-odd coupling to the time arrow. We describe both by an effective Lagrangian with a
U(1)-violating coupling $g_c\,\delta\mathcal O_S\Phi^2$ to the spacetime vacuum and a time bias $-\partial_\mu f(R)\,J^\mu$. For $f=c_a\ln|R|$ the chemical potential of producing a particle is
$\mu\simeq2c_a\dot H/H$. The asymmetry per pair reaches an attractor in the pair-dominated episode and freezes when the cosmological term is exhausted. The observed value requires
$|c_a|/\tau_S\simeq(0.65$--$1.0)\times10^{-4}$, with $T_S=\tau_SH/2\pi$ the temperature of the spacetime vacuum, and its sign is fixed by the arrow of time. The Yukawa decays of the pairs transfer the asymmetry, with a uniform sign, to baryons and leptons and to dark-matter particles. The current of the charged decay products generates a magnetic field of
$10^{-14}$--$10^{-11}$\,G today on the coherence length of reheating.
}


\begin{document}
\maketitle
\flushbottom

\section{Introduction}\label{sec:intro}

\paragraph{Inflation, reheating and observations.} Inflation was proposed to solve the horizon, flatness and monopole problems of the hot big bang~\cite{Starobinsky1980,Guth1981,Linde1982,AlbrechtSteinhardt1982}.
Its most important consequence is that quantum fluctuations, stretched beyond the Hubble radius, seed nearly scale-invariant adiabatic curvature perturbations and tensor modes
\cite{MukhanovChibisov1981,Hawking1982,Starobinsky1982,GuthPi1982,BardeenSteinhardtTurner1983}. The theory of these perturbations and of the many models of inflation is reviewed in
Refs.~\cite{MukhanovFeldmanBrandenberger1992,LythRiotto1999,Baumann2009,MartinRingevalVennin2014}.

After inflation, the energy that drove it must be converted into the hot plasma of the standard cosmology. Two broad pictures exist:
\begin{itemize}[leftmargin=1.5em,itemsep=1pt]
\item \emph{Perturbative reheating}, through the decay of the inflaton~\cite{AbbottFarhiWise1982,DolgovLinde1982};
\item \emph{Non-perturbative preheating}, through parametric resonance~\cite{TraschenBrandenberger1990,KofmanLindeStarobinsky1994,ShtanovTraschenBrandenberger1995,KofmanLindeStarobinsky1997}.
\end{itemize}
Both are reviewed in Refs.~\cite{BassettTsujikawaWands2006,Allahverdi2010,Amin2015}. The predictions of
inflation have been tested with great precision.
\begin{itemize}[leftmargin=1.5em,itemsep=1pt]
\item \emph{Scalar spectrum.} Planck measures $n_s=0.965\pm0.004$ and an amplitude $A_s\simeq2.1\times10^{-9}$, with no detectable running, non-Gaussianity or isocurvature~\cite{Planck2018VI,Planck2018X}.
\item \emph{Tensor modes.} BICEP/Keck bounds the tensor-to-scalar ratio to $r<0.036$~\cite{BICEPKeck2021}.
\item \emph{Reheating.} It is constrained indirectly: its temperature must exceed a few MeV for nucleosynthesis, and its duration enters the number of e-folds $N_*$ between the exit of the
pivot scale and the end of inflation.
\end{itemize}

\paragraph{Particle production in an expanding Universe.} A time-dependent background produces particles from the vacuum. In expanding universes this was first studied by
Parker~\cite{Parker1968,Parker1969} and by Zel'dovich and Starobinsky~\cite{ZeldovichStarobinsky1972}; see Refs.~\cite{BirrellDavies,Ford2021} for reviews. Two regimes are distinguished.
\begin{itemize}[leftmargin=1.5em,itemsep=1pt]
\item \emph{Adiabatic regime.} When the expansion is slow compared with the frequency of a mode, the mode evolves adiabatically. The vacuum is then polarised, and real particles are
produced only with an exponentially small rate.
\item \emph{Non-adiabatic regime.} When adiabaticity breaks down, through a rapid change of the background or an oscillation, real particles are produced efficiently. This is the
mechanism of preheating.
\end{itemize}
Horizons add a thermal aspect. The de~Sitter horizon radiates at the temperature $H/2\pi$~\cite{GibbonsHawking1977}, as black-hole horizons do at the Hawking
temperature~\cite{Hawking1975}. Particles are created in pairs, which carry no net charge unless the background or the interactions distinguish particles from antiparticles. Polarisation,
the breakdown of adiabaticity and the thermal character of horizons underlie the present article.

\paragraph{Matter--antimatter asymmetry.} The Universe contains matter and essentially no antimatter. The baryon asymmetry, measured independently by the CMB and by big-bang
nucleosynthesis, is $n_B/s\simeq8.7\times10^{-11}$~\cite{Planck2018VI,Fields2020}. Its dynamical generation requires baryon-number violation, C and CP violation, and a departure from
thermal equilibrium~\cite{Sakharov1967}. Many mechanisms satisfy these conditions: decays of heavy particles in grand unification, electroweak baryogenesis~\cite{Kuzmin1985},
leptogenesis~\cite{FukugitaYanagida1986}, the Affleck--Dine mechanism~\cite{AffleckDine1985}, and spontaneous baryogenesis driven by a rolling field coupled to a conserved
current~\cite{CohenKaplan1987,CohenKaplan1988}; see Refs.~\cite{RiottoTrodden1999,DineKusenko2004} for reviews. In each case the asymmetry is tied to the mechanism that reheats the
Universe, or to a field that evolves during or after inflation.

\paragraph{The present scenario.} In a series of papers~\cite{Xue2023,Xue2023b,Xue2024} a scenario was proposed in which a time-varying cosmological term produces massive
particle--antiparticle pairs from the spacetime vacuum.
\begin{itemize}[leftmargin=1.5em,itemsep=1pt]
\item \emph{Inflation.} The pair density $\rhoMH\propto H^2$ slows the expansion and determines the spectral index~\cite{Xue2023}. The model has no inflaton.
\item \emph{Reheating.} The pairs dominate, oscillate between creation and annihilation, and decay into radiation~\cite{Xue2023b}.
\item \emph{Late times.} The exchange between matter and the cosmological term bears on the coincidence problem~\cite{Xue2024}.
\end{itemize}
In these previous articles, we derived the cosmic rate equation of this scenario,
\begin{equation}
  \dot\rho_M+3H\rho_M=\GaM\big(\rhoMH-\rho_M\big)-\Gde\,\rho_M ,
  \label{eq:rate}
\end{equation}
and its closed sets of background equations from an explicit field theory~\cite{Xue2026b}. Here $\rhoMH=2\chi m^2H^2$ is the pair density set by the expansion, $\GaM$ the relaxation rate and
$\Gde$ the decay rate. Reference~\cite{XuePRD} then presented a complete analysis of the cosmic history with these equations. It connected the inflation of Ref.~\cite{Xue2023} and the reheating of Ref.~\cite{Xue2023b} into one
self-consistent evolution, in agreement with $n_s$, $A_s$ and $r$. It also showed, for the first time, that a pre-inflation with relativistic pairs explains the low-$\ell$ power anomaly of the CMB. Since
the pairs are produced and annihilated together and the relaxation conserves the charge, this pair production is U(1)-symmetric. Whether and how it can generate the observed
asymmetry is the subject of this article.

\paragraph{Issues addressed in this article.}
\begin{enumerate}[label=(\roman*),leftmargin=1.6em,itemsep=2pt]
\item \emph{Charge separation at horizons.} Real pairs are produced continuously, from pre-inflation to reheating, and their members are separated by horizons. How much charge
does this put into today's Hubble volume? Under which condition is it a volume effect rather than a surface effect?
\item \emph{Bias and effective description.} What supplies the sign of the asymmetry? How is the resulting U(1) violation described by an effective Lagrangian, and what values of its
parameters reproduce the observed $n_B/s$ along the cosmic history of Refs.~\cite{XuePRD,Xue2023b,Xue2024}, consistently with the observations?
\end{enumerate}

\begin{figure}[t]
\centering
\includegraphics[width=\linewidth]{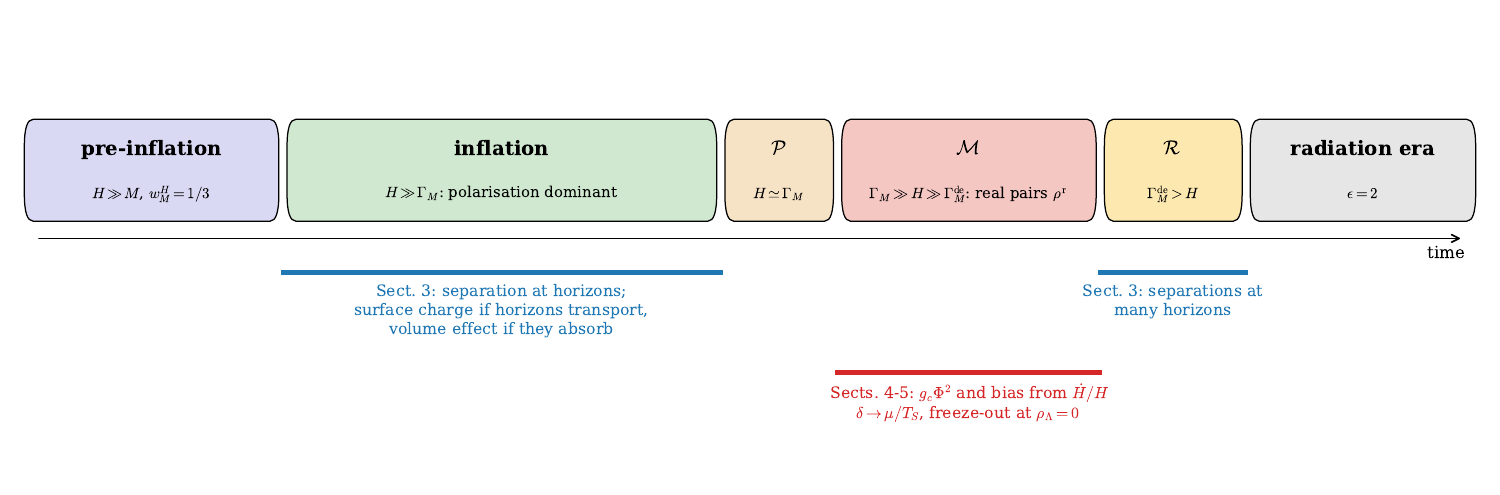}
\caption{Overview: the cosmic history of Refs.~\cite{XuePRD,Xue2023b,Xue2024} (Sect.~\ref{sec:field}), charge separation at horizons (Sect.~\ref{sec:horizons}), and charge-biased production in the
pair-dominated episode (Sects.~\ref{sec:effL}--\ref{sec:asymM}).}
\label{fig:timeline}
\end{figure}

\paragraph{Main results.}
\begin{itemize}[leftmargin=1.5em,itemsep=2pt]
\item \emph{Charge separation at horizons} (Sect.~\ref{sec:horizons}). Horizons absorb the charge they separate. Separation at the many horizons inside today's Hubble volume is then a volume
violation of the U(1) charge, at the rate $\Gamma_C=\kappa_C\GaM$ proportional to pair production.
\item \emph{Effective Lagrangian and the asymmetry} (Sects.~\ref{sec:effL}--\ref{sec:asymM}). A U(1)-violating coupling $g_c\,\delta\Ocal_S\Phi^2$ and a time bias $-\partial_\mu f(R)J^\mu$ give an attractor $\delta_{\rm eq}=\mu/T_S$ in the
pair-dominated episode. The observed value requires $|c_a|/\tau_S\simeq(0.65$--$1.0)\times10^{-4}$.
\item \emph{Baryogenesis, leptogenesis and darkogenesis} (Sect.~\ref{sec:transfer}). The Yukawa decays transfer the asymmetry to baryons and leptons, which survives the sphalerons when the decays
carry $B-L$, and to darkons, with an average darkon mass of about $1.8$\,GeV for equal visible and dark Yukawa couplings.
\item \emph{Magnetogenesis} (Sect.~\ref{sec:magneto}). The current of the charged decay products generates $B_0\simeq10^{-14}$--$10^{-11}$\,G today, on the coherence length of reheating.
\end{itemize}

\paragraph{Outline.} Section~\ref{sec:field} gives the unbiased Lagrangian and reviews pair production and the cosmic history. Section~\ref{sec:horizons} studies charge separation at horizons.
Section~\ref{sec:effL} introduces the effective Lagrangian and solves the charge equation along the cosmic history. Section~\ref{sec:asymM} computes the asymmetry of the pair-dominated episode and compares it
with observations. Sections~\ref{sec:transfer} and~\ref{sec:magneto} treat baryogenesis, leptogenesis, darkogenesis and magnetogenesis. Section~\ref{sec:conclusions} gives conclusions and remarks, and the
appendices collect derivations, the kinetic equations and the numerical methods.

\paragraph{Conventions.} We use natural units $\hbar=c=k_B=1$, the reduced Planck mass $\mpl=(8\pi G)^{-1/2}=2.435\times10^{18}$\,GeV, and the signature $(-,+,+,+)$. The
$\epsilon$-rate is $\epsilon=-\dot H/H^2$. The notation follows Ref.~\cite{Xue2026b}.

\section{Framework: unbiased Lagrangian, pair production and cosmic history}\label{sec:field}\label{sec:frw}

\subsection{Closed background system with a time-varying cosmological term}\label{sec:closedset}

The Friedmann equations with a time-varying cosmological term $\rho_\Lambda$, pairs $\rho_M$ and radiation $\rho_R$ [Ref.~\cite{Xue2026b}, Sect.~2, Eqs.~(2.3)--(2.4)], together with the cosmic rate equation
\eqref{eq:rate} and the radiation equation $\dot\rho_R+4H\rho_R=\Gde\rho_M$ [Eq.~(4.9)], form a closed system for $(H,\rho_\Lambda,\rho_M,\rho_R)$. The energy transfer from the cosmological term to
the matter is gravitational pair production, $\mathcal Q=-\dot\rho_\Lambda=\GaM(\rhoMH-\rho_M)$ [Eq.~(4.10)]. With $x=\ln(a/a_*)$, $h=H/H_*$ and $\Omega_i=\rho_i/(3\mpl^2H_*^2)$, the closed set reads
[Ref.~\cite{Xue2026b}, Sect.~5.1, Eq.~(5.1); Ref.~\cite{XuePRD}, Eqs.~(4.3)--(4.4)]
\begin{equation}
\begin{gathered}
  h^2=\Omega_\Lambda+\Omega_M+\Omega_R,\qquad \frac{\dd h^2}{\dd x}=-3\Omega_M-4\Omega_R,\\
  \frac{\dd\Omega_M}{\dd x}+3\Omega_M=\frac{\GaM}{H}\big(\Omega_M^H-\Omega_M\big)-\frac{\Gde}{H}\,\Omega_M,\qquad
  \frac{\dd\Omega_R}{\dd x}+4\Omega_R=\frac{\Gde}{H}\,\Omega_M,
\end{gathered}
\label{eq:closedset}
\end{equation}
with
\begin{equation}
  \Omega_M^H=\frac23\chi\frac{m^2}{\mpl^2}h^2,\qquad \frac{\GaM}{H}=\frac{\chi}{4\pi}\frac{m}{H_*}\frac{\epsilon}{h},\qquad \frac{\Gde}{H}=g_Y^2\frac{m}{H_*}\frac1h,\qquad
  \epsilon=\frac{\tfrac32\Omega_M+\tfrac43\Omega_R}{h^2}.
  \label{eq:closedaux}
\end{equation}
The two rates are [Ref.~\cite{XuePRD}, Eqs.~(4.5)--(4.6)]
\begin{equation}
  \GaM\simeq\frac{\chi m\epsilon}{4\pi},\qquad \Gde=g_Y^2m,\qquad g_Y^2=\frac{y^2}{8\pi}\frac{M}{m}.
  \label{eq:rates}
\end{equation}
The relaxation rate $\GaM$ is derived from the kinetic equation of the pairwise coupling to the spacetime vacuum [Ref.~\cite{Xue2026b}, Sect.~4.2, Eqs.~(4.3)--(4.5), and Sect.~5.4, Eq.~(5.4)];
its coefficient requires a microscopic theory of the spacetime vacuum.  With the mass function $m(x)$ of Sect.~\ref{sec:Leff}, Eqs.~\eqref{eq:closedset}--\eqref{eq:rates} describe the
whole history used in this article (Sect.~\ref{sec:transition}).

\subsection{Total Lagrangian, field equation, Noether current and modes}\label{sec:Leff}

The total Lagrangian of the framework is~\cite[Sect.~4.4, Eq.~(4.8)]{Xue2026b}
\begin{align}
  \mathcal L_{\rm eff}={}&\sqrt{-g}\,\Big[
  \underbrace{-|\partial\Phi|^2-M^2|\Phi|^2-\xi R|\Phi|^2-\delta\Ocal_S|\Phi|^2}_{\text{pairs and pairwise coupling to spacetime}}\nonumber\\
  &\qquad\underbrace{-\bar\psi\gamma^\mu\nabla_\mu\psi-\bar N\gamma^\mu\nabla_\mu N-y\left(\Phi\bar\psi N+{\rm h.c.}\right)}_{\text{decay}}\Big]
  +\mathcal L_{\mathcal S}.
  \label{eq:Leff}
\end{align}
It has three parts. The first describes the pairs by a complex scalar field $\Phi$ with non-minimal coupling $\xi$. The second is a pairwise coupling
$-\delta\Ocal_S|\Phi|^2$ to the fluctuations $\delta\Ocal_S$ of the spacetime vacuum, whose dynamics is $\mathcal L_{\mathcal S}$ \cite[Sect.~4.2 and Appendix~A]{Xue2026b}. The third is a Yukawa coupling to a light
fermion $\psi$ of U(1) charge $+1$ and a neutral fermion $N$. All terms have mass dimension 4 and are invariant under the global U(1) transformation $\Phi\to e^{i\alpha}\Phi$, $\psi\to e^{i\alpha}\psi$. The interactions are summarised in Sect.~\ref{sec:rates}.

\paragraph{Collective field, mass scale and degeneracy.} The field $\Phi$ is a collective quantum field: it represents $g_*$ degenerate pair states (particle species and their internal
states), each with the mass scale $M$ and each described by \eqref{eq:Leff}. Densities of the collective field are $g_*$ times those of a single state. For the polarisation density of
Sect.~\ref{sec:rates} this amounts to an effective mass
\begin{equation}
  m^2\equiv g_*M^2
  \label{eq:m_gmu}
\end{equation}
(for states of different masses, $m^2=\sum_fg_fM_f^2$, where $g_f$ is the degeneracy of the state $f$ with mass $M_f$). The two masses play different roles. The background depends only on $m$: through $\rhoMH=2\chi m^2H^2$, the $\epsilon$-rate $\epsilon=(1+w_M)\chi
m^2/\mpl^2$ and the relaxation rate $\GaM\simeq\chi m\epsilon/4\pi$ ($\chi\sim 10^{-3}$). The kinematics of the quanta depends on $M$: whether they are relativistic, the adiabaticity of the modes, and the suppression
of real-pair production. In particular $m$ can exceed the Planck mass while $M<H$.

\paragraph{The mass function $m(H)$.} Neither $M$ nor $g_*$ is determined within the present framework; both depend on the microscopic physics of the pairs. We therefore treat $m$ as a mass
function $m(H)$. Within a given dynamical epoch it is a constant, or more precisely a smooth, slowly varying function of $H$. It changes in the transition from one epoch to the next, where
the mass scale $M$ and the number $g_*$ of states change. Two such transitions occur
in this article, and in both the $\epsilon$-rate changes.
\begin{itemize}[leftmargin=1.5em,itemsep=2pt]
\item \emph{Pre-inflation $\to$ inflation.} Light pairs ($M\ll H$) are replaced by heavy ones ($M\gg H$): the pairs turn from relativistic to non-relativistic, the equation of state of the pair density drops from $1/3$ to $0$, and $m$ takes its inflationary value $m_*$~\cite{XuePRD,Xue2023b,Xue2024}.
\item \emph{Inflation $\to$ $\mathcal P$-episode.} $H$ falls below the relaxation rate $\GaM$. Adiabaticity breaks down, real pairs of heavier states and of more species are produced, so $M$ and $g_*$ grow and $m$
increases from $m_*$ to $m_{\mathcal M}$ (Sect.~\ref{sec:transition}).
\end{itemize}
The values of $m$ in the different epochs are fixed by observations.

\paragraph{Free field.} For one of the $g_*$ states the free part of \eqref{eq:Leff} is
\begin{equation}
  S_\Phi=\int\dd^4x\sqrt{-g}\,\Big[-g^{\mu\nu}\partial_\mu\Phi^*\partial_\nu\Phi
         -M^2|\Phi|^2-\xi R\,|\Phi|^2\Big],
  \label{eq:SPhi}
\end{equation}
where $\xi$ is the non-minimal coupling (a quartic term can be added; it only modifies the
pressure and is not needed below). Varying with respect to $\Phi^*$ gives, in the FLRW metric,
\begin{equation}
  \ddot\Phi+3H\dot\Phi-\frac{\nabla^2\Phi}{a^2}+\meff^2(t)\,\Phi=0,
  \qquad \meff^2\equiv M^2+\xi R.
  \label{eq:KG}
\end{equation}
The action is invariant under $\Phi\to e^{i\alpha}\Phi$. The conserved Noether current $J^\mu=-i(\Phi^*\partial^\mu\Phi-\Phi\,\partial^\mu\Phi^*)$, normalised so that particles carry charge $+1$,
has the time component and the physical spatial component (orthonormal frame, $J^{\hat\jmath}=aJ^j$)
\begin{equation}
  J^0=i\left(\Phi^*\dot\Phi-\Phi\dot\Phi^*\right),\qquad
  \JJ\equiv aJ^{j}\,\hat{\mathbf e}_j=\frac{2}{a}\,\mathrm{Im}\!\left(\Phi^*\nabla\Phi\right),
  \label{eq:J0J}
\end{equation}
and current conservation in FLRW reads
\begin{equation}
  \frac{1}{a^3}\frac{\partial}{\partial t}\!\left(a^3J^0\right)+\frac1a\nabla\cdot\JJ=0 .
  \label{eq:Jcons}
\end{equation}
A single positive-frequency wave, $\Phi=e^{-iMt}/\sqrt{2M}$, has $J^0=+1$, and a negative-frequency
wave, $\Phi^*=e^{+iMt}/\sqrt{2M}$, has $J^0=-1$.

\paragraph{Mode expansion and number densities.} We expand the field in modes,
\begin{equation*}
  \Phi(t,\xx)=\int\!\frac{\dd^3k}{(2\pi)^3}\left[a_\kk\,u_k(t)+b^\dagger_{-\kk}\,u_k^*(t)\right]e^{i\kk\cdot\xx},
\end{equation*}
with $[a_\kk,a^\dagger_{\kk'}]=[b_\kk,b^\dagger_{\kk'}]=(2\pi)^3\delta^{(3)}(\kk-\kk')$.
Setting $u_k=a^{-3/2}\chi_k$, Eq.~\eqref{eq:KG} becomes an oscillator equation,
\begin{gather}
  \ddot\chi_k+\Omega_k^2(t)\chi_k=0,\qquad
  \Omega_k^2=\frac{k^2}{a^2}+M^2+\Delta(t),\nonumber\\
  \Delta\equiv\xi R-\tfrac94H^2-\tfrac32\dot H
  =H^2\big[(12\xi-\tfrac94)-(6\xi-\tfrac32)\epsilon\big],
  \label{eq:Omega}
\end{gather}
with Wronskian normalisation $\chi_k\dot\chi_k^*-\chi_k^*\dot\chi_k=i$. For $M\gg H$ the adiabatic
solution is $\chi_k\simeq e^{-i\int\Omega_k\dd t}/\sqrt{2\Omega_k}$.

In a spatially homogeneous state with occupations
$\langle a^\dagger_\kk a_{\kk'}\rangle=(2\pi)^3\delta^{(3)}(\kk-\kk')f_\Phi(k)$ and likewise
$f_{\Phi^*}(k)$ for $b$, the physical number densities are
$n_{\Phi,\Phi^*}=a^{-3}\int\frac{\dd^3k}{(2\pi)^3}f_{\Phi,\Phi^*}(k)$. Using
$i(u_k^*\dot u_k-u_k\dot u_k^*)=i\,a^{-3}(\chi_k^*\dot\chi_k-\chi_k\dot\chi_k^*)=a^{-3}$ (the
$H$-terms cancel in the antisymmetric combination), the normal-ordered charge density is $\langle{:}J^0{:}\rangle=a^{-3}\int\frac{\dd^3k}{(2\pi)^3}[f_\Phi(k)-f_{\Phi^*}(k)]=n_\Phi-n_{\Phi^*}$,
where interference terms $\propto\langle a\,b\rangle e^{-2i\int\Omega}$ have been averaged over
times $\gg M^{-1}$. Similarly $\langle{:}\rho{:}\rangle=a^{-3}\int\frac{\dd^3k}{(2\pi)^3}\Omega_k(f_\Phi+f_{\Phi^*})
\to M(n_\Phi+n_{\Phi^*})$ for non-relativistic occupations, per state.

\subsection{Pair production}\label{sec:rates}\label{sec:closed}\label{sec:reservoir}\label{sec:pairwise}\label{sec:yukawa}\label{sec:coherent}

We summarise the results of Refs.~\cite{Xue2026b,XuePRD} that are used below.
\begin{itemize}[leftmargin=1.5em,itemsep=2pt]
\item \emph{Adiabatic expansion: polarisation.} When the background changes slowly compared with the frequency of the pairs, the modes evolve adiabatically. The expansion then
polarises the spacetime vacuum: it creates the pair density $\rhoMH=2\chi m^2H^2$. This density follows the
expansion instantaneously and disappears when the expansion stops. A smooth expansion produces real pairs only with the suppression $e^{-2\pi M/H}$.
\item \emph{Relaxation.} The pair density relaxes towards $\rhoMH$ through the pairwise coupling to the fluctuations $\delta\Ocal_S$ of the spacetime vacuum, at the rate $\GaM\simeq\chi m\epsilon/4\pi$
[Ref.~\cite{Xue2026b}, Sects.~4.2 and~5.4]. The kinetic equation is Eq.~(4.4) of Ref.~\cite{Xue2026b}. Particles and antiparticles are created and annihilated together, so it
conserves the charge. The Yukawa decays $\Phi\to\psi\bar N$, $\Phi^*\to\bar\psi N$ occur at the rate $\Gde$ of Eq.~(\ref{eq:rates}).
\item \emph{Cosmic rate equation and closed set.} The pair density obeys the rate equation \eqref{eq:rate}, which, with the Friedmann equations, forms the closed set
\eqref{eq:closedset}. For $\GaM\gg H$, the Hubble function acquires a fast component oscillating at the frequency $2M$, which produces pairs resonantly [Ref.~\cite{Xue2026b}, Sects.~6--8].
\item \emph{From polarisation to real pairs.} During pre-inflation and inflation, $\GaM<H$, the expansion is adiabatic for the pairs, and the pair density is the polarisation. When $H$ falls to $\GaM$,
adiabaticity breaks down [Ref.~\cite{XuePRD}, Sect.~4.4]. The fast oscillation of $H$ then breaks adiabaticity mode by mode. The polarisation turns into real pairs, which oscillate between
creation and annihilation at the frequency $2M$ and are produced resonantly, without the suppression $e^{-2\pi M/H}$.
\end{itemize}

\paragraph{Continuous pair production at the rate $\GaM$.} The gain term of the rate equation \eqref{eq:rate}, $\GaM\rhoMH$, is the continuous production of pairs from the spacetime vacuum, and the loss
term $\GaM\rho_M$ their annihilation into it. In terms of numbers, the pairs produced per unit volume and time are
\begin{equation}
  \dot n_{\rm prod}=\GaM\,n_M^H ,
  \label{eq:nprod}
\end{equation}
and the number produced in a Hubble volume per e-fold is $(\GaM/H)\,N_H$, with $N_H$ the number of pairs in it. Production never stops while the cosmological term lasts, but its rate differs
from epoch to epoch: $\GaM\ll H$ in pre-inflation, $\GaM/H=e^{\epsilon x}\le1$ during inflation, and $\GaM/H\sim10^4$--$10^7$ in the pair-dominated episode (Sect.~\ref{sec:transition}). Production stops
when the cosmological term is exhausted. During pre-inflation and inflation this production sustains the polarisation, and real pairs remain suppressed as $e^{-2\pi M/H}$, except at
horizons (Sect.~\ref{sec:adrelax}). After adiabaticity breaks down, the production is of real pairs, without suppression. Every pair is created with zero charge, $\Phi(\kk)\Phi^*(-\kk)$, so this production alone conserves the charge. The kinetic origin of $\GaM$ is
summarised in Appendix~\ref{app:kinetic}. In Sects.~\ref{sec:horizons} and~\ref{sec:effL}, every charge-changing event is tied to a pair-production event, so that the charge-production rate is a
fraction of $\GaM$, $\Gamma_C=\kappa_C\GaM$.

\paragraph{Real pairs at horizons.}\label{sec:adrelax}\label{sec:vacpol} Real particles arise in the thin layer of width $\lambda_m=(\chi m)^{-1}$ near the horizon, where the polarisation
crosses it. Let $\eta_r$ be the number of real pairs created at the horizon of one Hubble sphere per e-fold. Two estimates bracket it: (A) $\eta_r=2\pi\GaM/H=\epsilon\chi m/2H$, the rate of
Ref.~\cite{Xue2023} applied at the horizon; and (B) $\eta_r=n_M^H\times4\pi H^{-2}\times\lambda_m=4\pi$, the polarisation content of the horizon layer crossing once per e-fold. When today's
Hubble scale exits, (A) gives $\eta_r\simeq2$ and (B) $\eta_r\simeq13$. Every such real pair is born straddling the horizon of the Hubble sphere in which it is created.

\paragraph{Cosmic history.}\label{sec:transition}

\begin{figure}[t]
\centering
\includegraphics[width=\linewidth]{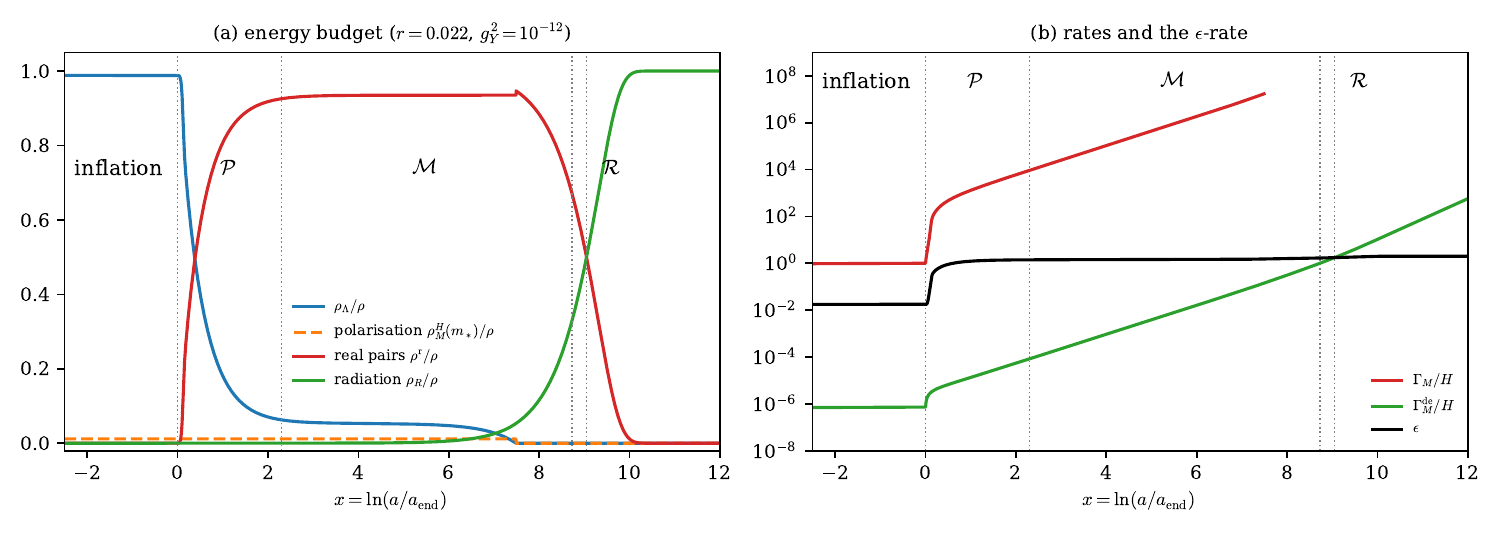}
\caption{Cosmic evolution from inflation to reheating from the closed set \eqref{eq:closedset}--\eqref{eq:rates}, for the baseline of Table~\ref{tab:baseline} (from Ref.~\cite{XuePRD}, Fig.~2; see also Refs.~\cite{Xue2023b,Xue2024}). (a) Energy fractions:
cosmological term, polarisation $\rhoMH(m_*)$, real pairs $\rho^{\rm r}$ and radiation. (b) Rates $\GaM/H$, $\Gde/H$ and $\epsilon$. Dotted lines separate inflation, the $\mathcal P$-, $\mathcal M$- and $\mathcal R$-episodes. Charge
production at the rate $\Gamma_C=\kappa_C\GaM$ follows the red curve in (b).}
\label{fig:evolution}
\end{figure}

The closed set \eqref{eq:closedset}--\eqref{eq:rates}, with the mass function $m(x)$, gives the self-consistent history of Refs.~\cite{XuePRD,Xue2023b,Xue2024}. It agrees with $n_s$, $A_s$, $r$, the horizon-matching $N_*$, the
reheating constraints and the low-$\ell$ power anomaly. Table~\ref{tab:history} and Fig.~\ref{fig:evolution} summarise it, and Table~\ref{tab:baseline} lists the baseline parameters.

\begin{table}[t]
\centering\small
\begin{tabular}{@{}p{2.1cm}p{2.4cm}cp{1.9cm}p{4.7cm}@{}}
\toprule
epoch & condition & $\epsilon$ & $\GaM/H$ & features\\
\midrule
pre-inflation & $\GaM\ll H$, $H\gg M$ & $0.019$--$0.47$ & $\ll1$ & relativistic pairs, $w_M^H=1/3$, $m_{\rm pre}\simeq(0.9$--$4.5)\,m_*$; ends $3.5$--$10$ e-folds before the pivot exit [Sect.~6, Eqs.~(6.1)--(6.2), Table~2]\\
inflation & $\GaM<H$ & $0.0175$ & $e^{\epsilon x}\to1$ & polarisation dominant; pivot exits $N_*\simeq53.5$, today's scale $N_0\simeq58.9$ e-folds before the end [Sect.~5.2]\\
$\mathcal P$-episode & $H\simeq\GaM$ & $0.0175\to1.42$ & $1\to10^4$ & breakdown of adiabaticity, real pairs; $m_*\to m_{\mathcal M}$ [Eq.~(5.2)]; $0\le x\le2.3$ [Sect.~5.3]\\
$\mathcal M$-episode & $\GaM\gg H\gg\Gde$ & $1.42$ & $10^4$--$10^7$ & real pairs in detailed balance; $\rho_\Lambda$ exhausted at $x_\Lambda=7.49$ [Sect.~5.4]\\
$\mathcal R$-episode & $\Gde>H$ & $\to2$ & --- & decays; radiation dominates from $x=9.05$, $T_{\rm RH}\simeq6\times10^{12}$\,GeV [Sect.~5.5]\\
\bottomrule
\end{tabular}
\caption{Epochs of the cosmic history of Refs.~\cite{XuePRD,Xue2023b,Xue2024}, from the closed set \eqref{eq:closedset}--\eqref{eq:rates}; $x=\ln(a/a_{\rm end})$. Section and equation numbers refer to Ref.~\cite{XuePRD}; see also Refs.~\cite{Xue2023b,Xue2024}.}
\label{tab:history}
\end{table}

\begin{table}[t]
\centering\small
\begin{tabular}{@{}p{2.4cm}p{12.2cm}@{}}
\toprule
inflation & $\epsilon=0.0175$, $\chi=9.9\times10^{-4}$, $m_*=4.21\,\mpl$, $H_*=1.48\times10^{-5}\,\mpl$, $H_{\rm end}=\GaM=5.80\times10^{-6}\,\mpl$ ($r=0.022$)\\
transition & $m^2:m_*^2\to m_{\mathcal M}^2$ over $\Delta x\simeq0.5$\\
$\mathcal M$-episode & $m_{\mathcal M}=37.9\,\mpl$ ($\epsilon_{\mathcal M}=1.42$), $\GaM/H\sim10^4$--$10^7$, $x_\Lambda=7.49$\\
decay & $g_Y^2=10^{-12}$ ($10^{-14}$--$10^{-10}$ for a pronounced $\mathcal M$-episode)\\
reheating & $T_{\rm RH}=6.2\times10^{12}$\,GeV ($6\times10^{11}$--$6\times10^{13}$\,GeV) [Sect.~5.5]\\
pre-inflation & $m_{\rm pre}\simeq(0.9$--$4.5)\,m_*$, $N_{\rm tr}=-3.5$ to $-10$, $M_{\rm pre}\lesssim4\times10^{12}$\,GeV, $g_{*,{\rm pre}}\gtrsim10^{13}$ [Sect.~6]\\
\bottomrule
\end{tabular}
\caption{Baseline parameters of the cosmic history of Ref.~\cite{XuePRD} [Sect.~5.6]; see also Refs.~\cite{Xue2023b,Xue2024}. Section numbers refer to Ref.~\cite{XuePRD}.}
\label{tab:baseline}
\end{table}

\section{Charge separation at horizons}\label{sec:horizons}\label{sec:unbiased}

With the U(1)-invariant Lagrangian \eqref{eq:Leff}, every pair $\Phi(\kk)\Phi^*(-\kk)$ is created with zero charge, and the relaxation and the decays conserve the charge (Sect.~\ref{sec:rates}). A local
charge can arise only by separating the members of pairs, and horizons are where this happens.

\subsection{Charge separation on a horizon}\label{sec:onhorizon}

Real pairs are created in the layer of width $\lambda_m=(\chi m)^{-1}$ at the horizon of each Hubble volume (Sect.~\ref{sec:adrelax}). A pair born there straddles the horizon: one member is inside, the
other outside. For an observer inside, the charge of the Hubble volume changes by $\pm1$ at each such event. The partner that is no longer inside has one of two fates.
\begin{enumerate}[label=(\Roman*),leftmargin=1.6em,itemsep=2pt]
\item \emph{It leaves the causal horizon.} The partner is causally disconnected from the inside for as long as the horizon persists. During inflation $H$ is nearly constant, the horizon
is an event horizon, and the partner does not return while inflation lasts. After inflation the comoving Hubble radius grows and the partner re-enters, but it is still somewhere
in the Universe, and the charge is conserved globally.
\item \emph{It is absorbed by the horizon.} The horizon is part of the spacetime vacuum, with which the pairs exchange energy at the rate $\GaM$. If the horizon also takes up the charge of
the partner, the charge is transferred to the spacetime vacuum, and the matter charge is not conserved. This is expected, since horizons do not conserve global charges (Sect.~\ref{sec:absorption}).
\end{enumerate}
Which member stays inside is random unless something distinguishes particles from antiparticles. A single Hubble patch at the exit of today's scale contains
$N_{\rm st}=\eta_r\times5.5\simeq11$--$72$ straddling pairs (estimates (A) and (B) of Sect.~\ref{sec:adrelax}), so its charge is a sum of random signs,
\begin{equation}
  Q_{\rm exit}\simeq\pm N_{\rm st}^{1/2}\simeq\pm(3\text{--}8).
  \label{eq:Qexit}
\end{equation}
The distinction between (I) and (II), and the bias of the sign, decide whether this mechanism can produce the observed asymmetry.

\subsection{Charge absorption at a horizon}\label{sec:absorption}

Fate (II) is the decisive one, and it is supported by a body of results on horizons and global charges.
\begin{itemize}[leftmargin=1.5em,itemsep=2pt]
\item \emph{Black holes do not carry global charges.} A static black hole is characterised by its mass, angular momentum and gauge charges only. Baryon number, or any charge without a
long-range gauge field, cannot be measured from outside: the no-hair property~\cite{Bekenstein1972a,Bekenstein1972b}. A black hole formed from baryons is indistinguishable from one formed
from antibaryons.
\item \emph{The horizon returns the energy, but not the charge.} A black hole emits thermal radiation at its Hawking temperature~\cite{Hawking1975}. The radiation is fixed by the
temperature and the gauge charges, and carries no memory of the global charge that fell in. Baryons that fall in are therefore converted into radiation with zero baryon number: the
global charge is destroyed, a process Zel'dovich called the gravitational annihilation of baryons~\cite{Zeldovich1976}.
\item \emph{Cosmological horizons behave in the same way.} The event horizon of de~Sitter space has a temperature $H/2\pi$ and an entropy, like a black-hole horizon~\cite{GibbonsHawking1977}. What
crosses it is lost to the observer inside, and what the horizon emits back is thermal radiation at $H/2\pi$, again without memory of the global charge.
\item \emph{Gravity does not respect global symmetries.} Global charges are expected to be violated by gravitational effects in general~\cite{Kallosh1995,BanksSeiberg2011}. Gauge charges, in
contrast, are conserved because their flux through any surface surrounding the horizon is measurable.
\end{itemize}

\paragraph{Absorption equals leaving for good.} For the observer inside a horizon, a charge that crosses it can be in only two states. Either it stays beyond the horizon forever, or it is
taken up by the horizon and re-emitted as thermal radiation without charge. In both cases it never returns as charge, and the charge accessible inside changes permanently, by $\pm1$ per
straddling pair. The two descriptions of fate (II), absorption and leaving for good, are therefore equivalent. They differ from fate (I) in one respect only: in (I) the partner returns
with its charge once the horizon recedes, so the charge is conserved within any region larger than the horizon. The
particle--antiparticle asymmetry from oscillation modes that cross the horizon and never return was studied in this sense in Ref.~\cite{Xue2025}.

\paragraph{When is a horizon an event horizon?} During inflation $H$ is nearly constant, and each comoving observer has an event horizon at the physical distance $H^{-1}$. A partner that crosses it
cannot return while inflation lasts. In the pair-dominated $\mathcal M$-episode, $\epsilon\simeq1.42>1$, the comoving Hubble radius grows and the horizon is not an event horizon in the strict sense. A pair born
straddling it is, however, separated faster than it can be re-joined: the horizon layer of width $\lambda_m$ exchanges pairs with the spacetime vacuum at the rate $\GaM\gg H$ (Sect.~\ref{sec:rates}). The
partner beyond the horizon is thus taken up into the reservoir long before the horizon recedes. In the present framework the spacetime vacuum is that reservoir, with the temperature
$T_S$ of Sect.~\ref{sec:kms}. The absorption of charge by horizons is then the U(1)-violating part of the coupling of the pairs to the spacetime vacuum, described effectively by the term $g_c\Phi^2$
of Sect.~\ref{sec:effL}.

\subsection{Transport only: a surface charge}\label{sec:singlepatch}

Today's Hubble volume is a fixed comoving region $V_0$ of radius $R_0$. It was a single Hubble patch when today's scale exited, $N_0\simeq58.9$ e-folds before the end of inflation. We observe only this
patch; charges in other patches need not be averaged with ours. If horizons only transport charge (case (I)), the U(1) current is conserved locally, and the charge of $V_0$ changes only
through its comoving boundary,
\begin{equation}
  \frac{\dd Q_0}{\dd t}=-\oint_{\partial V_0}\JJ\cdot\dd\mathbf A .
  \label{eq:gauss}
\end{equation}
A partner separated at the horizon of a smaller Hubble volume inside $V_0$ remains in $V_0$ and is observed today, so internal separations cancel. Only pairs straddling $\partial V_0$ count.
Their number, summed over inflation (Appendix~\ref{app:boundary}), is
\begin{equation}
  N_\partial\simeq\frac{4\eta_r}{2(1-\epsilon)}\,e^{2(1-\epsilon)N_0}\simeq10^{50.9}\text{--}10^{51.7}.
  \label{eq:Nboundary}
\end{equation}
The resulting charge, $|Q_0|\simeq N_\partial^{1/2}\simeq10^{25.4}$--$10^{25.8}$ for random signs and at most $N_\partial\simeq10^{51}$--$10^{52}$ for a coherent sign, is a surface charge. It falls short of the
observed baryon number of today's Hubble volume,
\begin{equation}
  Q_{\rm req}=\frac{n_B}{s}\,s_0V_0\simeq2.6\times10^{78},
  \label{eq:Qreq}
\end{equation}
by $10^{53}$ and $10^{27}$ respectively. The reason is geometric: the observed charge scales with the volume, $(R_0/r_H)^3$, and a boundary charge at most with the area, $(R_0/r_H)^2$, with
$R_0/r_H\simeq10^{25.1}$ at the end of inflation. Equivalently, the asymmetry per pair frozen at exit is diluted by the subsequent unbiased production,
\begin{equation}
  \delta_{\rm relic}\lesssim\delta_{\rm exit}\,e^{-(2-3\epsilon/2)N_0}\simeq10^{-52}.
  \label{eq:dilution}
\end{equation}
A charge separated once, or only at the boundary, is therefore not enough. The observed charge must be produced throughout $V_0$, continuously and with a fixed sign.

\subsection{Continuous charge production at many horizons}\label{sec:manyhorizons}

This is what happens in case (II). Four features define it.
\begin{enumerate}[label=(\roman*),leftmargin=1.6em,itemsep=2pt]
\item \emph{Continuous.} Charge separation accompanies the production of real pairs, Eq.~\eqref{eq:nprod}, continuously from pre-inflation to reheating, not once at exit.
\item \emph{At many horizons.} It occurs at the horizons of all Hubble volumes inside $V_0$, so it is a volume effect. The number of separations in $V_0$ is proportional to the number of pairs
produced in $V_0$.
\item \emph{Charge production proportional to pair production.} A fraction $\kappa_C$ of the production events straddle a horizon. In a Hubble volume of radius $H^{-1}$, the fraction of the volume
within $\lambda_m$ of the horizon is $3\lambda_mH$ (Appendix~\ref{app:boundary}), so
\begin{equation}
  \Gamma_C=\kappa_C\,\GaM,\qquad \kappa_C\simeq3\lambda_mH=\frac{3H}{\chi m},\qquad \kappa_C\frac{\GaM}{H}=\frac{3\epsilon}{4\pi}.
  \label{eq:k2geo}
\end{equation}
During inflation $\kappa_C\simeq4\times10^{-3}$ and $\kappa_C\GaM/H\simeq0.004$: separations are rare. In the $\mathcal M$-episode, $\kappa_C$ falls from $10^{-4}$ to $2\times10^{-8}$, but $\kappa_C\GaM/H\simeq0.34$ stays
constant, because both are proportional to $m$.
\item \emph{Biased in time.} Horizons are not static: they evolve with $\dot H/H=-\epsilon H$. An evolving horizon distinguishes the direction of time and can favour particles. We parameterise
the bias of the separations by
\begin{equation}
  \delta_{\rm eq}=-b_0\,\epsilon ,
  \label{eq:b0}
\end{equation}
proportional to the rate of change of the horizon.
\end{enumerate}
The net effect is an effective violation of the U(1) charge throughout $V_0$, at the rate $\Gamma_C$ and towards the biased value $\delta_{\rm eq}$. Its effective description is given in Sect.~\ref{sec:effL},
where Eq.~\eqref{eq:b0} becomes the scale-free bias $\mu\simeq2c_a\dot H/H$, with $b_0=4\pi c_a/\tau_S$.

\subsection{Bias in time and chemical potential}\label{sec:horizonbias}

\paragraph{The time arrow alone does not bias the charge.} The unbiased Lagrangian \eqref{eq:Leff} is invariant under charge conjugation C, $\Phi\leftrightarrow\Phi^*$, and so is the Friedmann background, which
couples to $\Phi$ and $\Phi^*$ identically. A process driven by the background therefore produces a particle $\Phi$ and an antiparticle $\Phi^*$ at the same rate. This holds for any history $H(t)$:
at a static or at an expanding horizon, and whether $H$ changes slowly or oscillates fast. The expansion does break time reversal T, through $\dot H<0$, and with it CPT. It does not break
C or CP, and Sakharov's conditions require an interaction that is odd under both~\cite{Sakharov1967}. An expanding horizon therefore separates particles and antiparticles in equal
numbers on average. To favour particles, the pairs need an interaction that distinguishes $\Phi$ from $\Phi^*$ and is tied to the time arrow.

\paragraph{A coupling of the current to the time arrow.} The minimal such interaction is a derivative coupling of the U(1) Noether current $J^\mu$ [Eq.~\eqref{eq:J0J}] to a scalar function $f$,
$-\partial_\mu f\,J^\mu$. It is C- and CP-odd, because $J^\mu$ is. In spontaneous baryogenesis $f$ is a rolling pseudoscalar field $\theta$~\cite{CohenKaplan1987,CohenKaplan1988}. If instead $f=f(R)$ is built from the
geometry, the coupling needs no new field~\cite{Davoudiasl2004}. In the Friedmann background it reads $-\dot f\,J^0$. With $J^0=n_\Phi-n_{\Phi^*}$, it lowers the energy needed to produce a particle and
raises that needed to produce an antiparticle, by
\begin{equation}
  \mu=\dot f(R).
  \label{eq:mu_intro}
\end{equation}
This is the chemical potential of producing a particle relative to its antiparticle. It is set by the rate of change of the expansion, i.e.\ by the time arrow.

\paragraph{The bias.} In equilibrium with the spacetime vacuum at temperature $T_S$ (Sect.~\ref{sec:kms}), particles are then favoured by $\delta_{\rm eq}=\mu/T_S$ per pair. This is the quantitative form of
the bias \eqref{eq:b0}, with a sign fixed by $\dot H<0$. Section~\ref{sec:effL} introduces the coupling in an effective Lagrangian, together with the U(1)-violating term that describes the
absorption of charge by horizons.

\section{Effective Lagrangian: U(1) violation and a bias from the expansion}\label{sec:effL}\label{sec:newL}\label{sec:gravbias}

\subsection{Requirements}

Section~\ref{sec:horizons} fixes the requirements for an effective description.
\begin{enumerate}[label=(R\arabic*),leftmargin=2.2em,itemsep=2pt]
\item \emph{Violation of the U(1) charge in the bulk.} The charge of today's Hubble volume must grow with its volume, which requires the matter charge to be violated locally
(Sect.~\ref{sec:manyhorizons}).
\item \emph{A bias of fixed sign.} A C- and CP-odd coupling to the time arrow (Sect.~\ref{sec:horizonbias}).
\item \emph{An attractor.} The asymmetry per pair must tend to a value independent of the initial state and of the number of e-folds.
\item \emph{Freeze-out.} The U(1) violation must stop before the bias disappears, so that the asymmetry is not washed out.
\item \emph{An unchanged background.} The tracking that gives $\epsilon$, $n_s$ and $r$, and the cosmic history of Refs.~\cite{XuePRD,Xue2023b,Xue2024}, must not be modified.
\end{enumerate}

\subsection{The effective Lagrangian}

We add to the unbiased Lagrangian \eqref{eq:Leff} the terms
\begin{equation}
  \Delta\mathcal L=\sqrt{-g}\,\Big[-\frac{g_c}{2}\,\delta\Ocal_S\big(\Phi^2+\Phi^{*2}\big)-\big(\partial_\mu f(R)\big)\,J^\mu\Big],
  \label{eq:DeltaL}
\end{equation}
\begin{itemize}[leftmargin=1.5em,itemsep=2pt]
\item \emph{The first term} couples $\Phi^2$ and $\Phi^{*2}$ to the same fluctuation operator $\delta\Ocal_S$ of the spacetime vacuum as the pairwise coupling [\cite{Xue2026b}, Eq.~(4.3)]. Since $\Phi^2$
carries charge $-2$, it creates and annihilates same-sign pairs, $\Phi\Phi$ and $\Phi^*\Phi^*$, and violates the U(1) charge by two units (R1). It is the effective description of the absorption
of charge by horizons (Sect.~\ref{sec:manyhorizons}). Its rates are $\Gamma_C^{\uparrow,\downarrow}=g_c^2S_{\mathcal O}(\mp2\Omega)/4\Omega^2$, as for the pairwise coupling, and the resulting rate is a fraction of
the pair rate,
\begin{equation}
  \Gamma_C=\kappa_C\,\GaM ,\qquad \kappa_C\propto g_c^2 .
  \label{eq:G2kappa}
\end{equation}
\item \emph{The second term} couples the U(1) Noether current $J^\mu$ of Sect.~\ref{sec:Leff} [Eq.~\eqref{eq:J0J}] to the gradient of a function of the Ricci scalar. This is the coupling of gravitational baryogenesis~\cite{Davoudiasl2004}, a
geometric form of spontaneous baryogenesis~\cite{CohenKaplan1987,CohenKaplan1988}. It is C- and CP-odd. In the Friedmann background it reads $-\dot f\,J^0$, which shifts the energies of
particles and antiparticles by $\pm\dot f$: a chemical potential $\mu=\dot f$ (R2).
\end{itemize}
The two terms are useful only together. The field redefinition $\Phi\to e^{if}\Phi$ removes the derivative coupling, up to terms of order $(\partial f)^2|\Phi|^2$, and moves $f$ into the phase of
the first term, $\Phi^2\to e^{2if}\Phi^2$. Without U(1) violation the chemical potential is therefore unphysical~\cite{DineHuetSingletonSusskind1991}; with it, a time-dependent CP-violating phase
biases the production of $\Phi\Phi$ against $\Phi^*\Phi^*$.

\paragraph{The U(1)-violating term in detail.} In terms of modes, $\Phi=\int_{\kk}[a_\kk u_k+b^\dagger_{-\kk}u_k^*]e^{i\kk\cdot\xx}$ (Sect.~\ref{sec:Leff}), so the two couplings to $\delta\Ocal_S$ contain different operators:
\begin{equation}
  |\Phi|^2\ni a_\kk b_{-\kk},\ a^\dagger_\kk b^\dagger_{-\kk},\qquad\qquad
  \Phi^2+\Phi^{*2}\ni a_\kk a_{-\kk},\ a^\dagger_\kk a^\dagger_{-\kk},\ b_\kk b_{-\kk},\ b^\dagger_\kk b^\dagger_{-\kk},
  \label{eq:modes_ops}
\end{equation}
together with number-conserving terms that only shift the frequencies. The pairwise coupling creates and annihilates particle--antiparticle pairs, $\Phi\Phi^*$, with zero charge. The new term
creates and annihilates same-sign pairs, $\Phi\Phi$ and $\Phi^*\Phi^*$, and changes the charge by $\pm2$. Both draw the energy $2\Omega_k$ from, or return it to, the spacetime vacuum. To lowest order in the
couplings, the golden-rule rates are fixed by the spectral function $S_{\mathcal O}(\omega)$ of $\delta\Ocal_S$ at $\omega=\mp2\Omega_k$ (Appendix~\ref{app:kinetic}):
\begin{equation}
  \Gamma^{\uparrow,\downarrow}_{M,k}=\frac{S_{\mathcal O}(\mp2\Omega_k)}{4\Omega_k^2}\quad(\Phi\Phi^*),\qquad\qquad
  \Gamma^{\uparrow,\downarrow}_{C,k}=g_c^2\,\frac{S_{\mathcal O}(\mp2\Omega_k)}{4\Omega_k^2}\quad(\Phi\Phi,\ \Phi^*\Phi^*).
  \label{eq:golden}
\end{equation}
The macroscopic rates follow by integrating over the modes. The pairwise channel relaxes the pair density towards $\rhoMH$ at the rate $\GaM$ [Ref.~\cite{Xue2026b}, Sects.~4.2 and~5.4]. The
same-sign channel relaxes the charge towards the biased value at the rate $\Gamma_C$, Eq.~\eqref{eq:G2kappa}. Since both are set by the same spectral function, $\kappa_C=\Gamma_C/\GaM$ measures the
relative strength of the two channels. In the horizon picture of Sect.~\ref{sec:manyhorizons} it is the fraction $3\lambda_mH$ of production events that straddle a horizon.

\begin{table}[t]
\centering\small
\begin{tabular}{@{}p{3.2cm}p{5.4cm}p{5.4cm}@{}}
\toprule
 & pair rate $\GaM$ & charge rate $\Gamma_C$\\
\midrule
coupling & $-\delta\Ocal_S|\Phi|^2$ (U(1)-invariant) & $-\tfrac{g_c}{2}\delta\Ocal_S(\Phi^2+\Phi^{*2})$ (U(1)-violating)\\
process & $\Phi\Phi^*\leftrightarrow$ spacetime vacuum & $\Phi\Phi\leftrightarrow$ vacuum, $\Phi^*\Phi^*\leftrightarrow$ vacuum\\
charge change & $0$ & $\pm2$\\
detailed balance & $\Gamma_{M,k}^\downarrow/\Gamma_{M,k}^\uparrow=e^{2\Omega_k/T_S}$ & $\Gamma_{C,k}^\downarrow/\Gamma_{C,k}^\uparrow=e^{2(\Omega_k\mp\mu)/T_S}$, biased by $\mu$\\
relaxes & $\rho_M\to\rhoMH$ & $n_Q\to\delta_{\rm eq}n_M$, $\delta_{\rm eq}=\mu/T_S$\\
macroscopic rate & $\GaM\simeq\chi m\epsilon/4\pi$, Eq.~\eqref{eq:rates} & $\Gamma_C=\kappa_C\GaM$, $\kappa_C$ from Eq.~\eqref{eq:k2geo}\\
during inflation & $\GaM/H=e^{\epsilon x}\le1$ & $\Gamma_C/H\simeq0.004$, Eq.~\eqref{eq:k2geo}\\
$\mathcal M$-episode & $\GaM/H\sim10^4$--$10^7$ & $\Gamma_C/H\simeq0.34$, Eq.~\eqref{eq:k2geo}; $\gg1$ for constant $\kappa_C\gtrsim10^{-4}$\\
switches off & when $\rho_\Lambda$ is exhausted & with $\GaM$, at $x_\Lambda=7.49$\\
effect on the expansion & sets $\rho_M$, which enters the closed set \eqref{eq:closedset} & none: it changes only the charge, $\delta\ll1$, not $\rho_M$\\
\bottomrule
\end{tabular}
\caption{Pair production and charge production compared. Both come from the coupling of the pairs to the fluctuations $\delta\Ocal_S$ of the spacetime vacuum; they differ in the charge
of the pairs they create and in the quantity they relax.}
\label{tab:GMG2}
\end{table}

The two rates are thus two channels of the same coupling to the spacetime vacuum (Table~\ref{tab:GMG2}). $\GaM$ creates neutral pairs, sustains the pair density, and enters the background.
$\Gamma_C$ creates charged pairs, carries the bias, and leaves the background unchanged. Charge production is tied to pair production, and it stops with it.

\paragraph{The bias from $\dot H/H$.} In the Friedmann background $R=6H^2(2-\epsilon)$, and
\begin{equation}
  \mu=\dot f(R)=f'(R)\,\dot R,\qquad \frac{\dot R}{HR}=-2\epsilon-\frac{\epsilon'}{2-\epsilon},\qquad \epsilon'=\frac{\dd\epsilon}{\dd x}.
  \label{eq:mu_R}
\end{equation}
The function $f(R)$ is not fixed by the framework, and other forms of the time bias are possible. For illustration we use the scale-free choice
\begin{equation}
  f=c_a\ln|R|,\qquad \mu=c_a\frac{\dot R}{R}\simeq2c_a\frac{\dot H}{H}=-2c_a\epsilon H ,
  \label{eq:scalefree}
\end{equation}
for slowly varying $\epsilon$. The bias is then set directly by the slow component $\dot H/H$ of the expansion, and its strength by a single dimensionless coefficient $c_a$.

With the temperature of the spacetime vacuum written as $T_S=\tau_S\,H/2\pi$ ($\tau_S=1$ for the Gibbons--Hawking value), the attractor of Sect.~\ref{sec:newL} becomes
\begin{equation}
  \delta_{\rm eq}=\frac{\mu}{T_S}=\frac{2\pi c_a}{\tau_S}\,\frac{\dot R}{HR}=-\frac{2\pi c_a}{\tau_S}\Big(2\epsilon+\frac{\epsilon'}{2-\epsilon}\Big).
  \label{eq:deq_grav}
\end{equation}
It is constant within each epoch: $\delta_{\rm eq}\simeq-4\pi\epsilon\,c_a/\tau_S=-0.22\,c_a/\tau_S$ during inflation and $-17.9\,c_a/\tau_S$ in the $\mathcal M$-episode ($\epsilon\simeq1.42$).

\subsection{Equilibrium, detailed balance and the temperature of the spacetime vacuum}\label{sec:kms}

The bias $\delta_{\rm eq}=\mu/T_S$ rests on three notions, which we make precise here.

\paragraph{The spacetime vacuum as a reservoir.} The pairs couple to the fluctuations $\delta\Ocal_S$ of the spacetime vacuum, whose dynamics is $\mathcal L_S$ [Ref.~\cite{Xue2026b}, Sect.~4.2 and Appendix~A]. For the pairs,
the spacetime vacuum is a reservoir: its state is not changed appreciably by the exchange of pairs, and it is characterised by the two-point function of $\delta\Ocal_S$. The rates of all
processes in which the pairs exchange energy $\omega$ with it are fixed by the spectral function $S_{\mathcal O}(\omega)=\int\dd t\,e^{i\omega t}\langle\delta\Ocal_S(t)\,\delta\Ocal_S(0)\rangle$: absorption from the reservoir at
$\omega<0$, emission into it at $\omega>0$ [Eq.~\eqref{eq:golden}].

\paragraph{KMS condition and temperature.} A state of the reservoir is an equilibrium state at temperature $T_S$ if its correlation functions satisfy the Kubo--Martin--Schwinger (KMS)
condition~\cite{Kubo1957,MartinSchwinger1959,HaagHugenholtzWinnink1967},
\begin{equation}
  \langle\delta\Ocal_S(t)\,\delta\Ocal_S(0)\rangle=\langle\delta\Ocal_S(0)\,\delta\Ocal_S(t+i/T_S)\rangle
  \qquad\Longleftrightarrow\qquad
  S_{\mathcal O}(+\omega)=e^{\omega/T_S}\,S_{\mathcal O}(-\omega).
  \label{eq:KMS}
\end{equation}
For a Gibbs state this is equivalent to the thermal density matrix $\propto e^{-H/T_S}$. The KMS form, however, does not require a Hamiltonian with a well-defined Gibbs state, and it applies to
the vacuum restricted to a region bounded by a horizon. The vacuum restricted to a Rindler wedge satisfies the KMS condition at the Unruh temperature with respect to boosts
(Bisognano--Wichmann theorem)~\cite{BisognanoWichmann1976,Unruh1976}. The de~Sitter vacuum restricted to a static patch satisfies it at the Gibbons--Hawking temperature $H/2\pi$ with
respect to the time translations of that patch~\cite{GibbonsHawking1977}; see Ref.~\cite{BirrellDavies} for a review. In both cases the temperature comes from the horizon. This motivates our
parameterisation $T_S=\tau_S\,H/2\pi$, with $\tau_S$ of order one if the spacetime vacuum seen by the pairs inherits the temperature of the expanding horizon. The value of $\tau_S$ for the
fluctuations $\delta\Ocal_S$ requires a microscopic theory of $\mathcal L_S$.

\paragraph{Detailed balance.} With the golden-rule rates \eqref{eq:golden}, the KMS condition gives the ratio of the emission and absorption rates of each channel,
\begin{equation}
  \frac{\Gamma_{M,k}^\downarrow}{\Gamma_{M,k}^\uparrow}=\frac{\Gamma_{C,k}^\downarrow}{\Gamma_{C,k}^\uparrow}=e^{2\Omega_k/T_S}.
  \label{eq:detbal}
\end{equation}
This is detailed balance: in equilibrium, each process and its inverse occur at equal rates. The pair occupations relax to the Bose--Einstein form $f=[e^{\Omega/T_S}-1]^{-1}$ with the temperature
of the reservoir. With the chemical potential $\mu$, the energies of particles and antiparticles are $\Omega\mp\mu$, and detailed balance gives $f_{\Phi,\Phi^*}=[e^{(\Omega\mp\mu)/T_S}-1]^{-1}$. Particles are then
favoured, by $\delta_{\rm eq}\simeq\mu/T_S$ per pair.

\paragraph{Equilibrium and the departure from it.} Two remarks delimit the use of these notions.
\begin{enumerate}[label=(\roman*),leftmargin=1.6em,itemsep=2pt]
\item \emph{Equilibrium is local in time.} The expansion is not stationary, so the KMS condition can hold only adiabatically, on time scales short compared with $H^{-1}$. The rates
$\GaM,\Gamma_C\gg H$ of the $\mathcal M$-episode ensure that the pairs equilibrate with the reservoir faster than the expansion changes it. During inflation $\Gamma_C<H$, and the charge does not reach
equilibrium.
\item \emph{No departure from equilibrium is needed for the bias.} Sakharov's third condition requires a departure from thermal equilibrium only if CPT holds. Here the expanding
background breaks CPT, and $\mu=\dot f(R)$ biases the equilibrium itself, as in spontaneous baryogenesis~\cite{CohenKaplan1987,CohenKaplan1988}. The departure from equilibrium enters only at
freeze-out, when $\Gamma_C$ falls below $H$ at the exhaustion of the cosmological term, and the asymmetry is frozen at its equilibrium value.
\end{enumerate}
The bias requires a finite temperature $T_S$ of the spacetime vacuum, i.e.\ an asymmetric spectrum $S_{\mathcal O}(+\omega)>S_{\mathcal O}(-\omega)$.

\subsection{Charge equation and solution along the cosmic history}\label{sec:solution}

\paragraph{Kinetic equations.} The occupations $f_\Phi(k)$ and $f_{\Phi^*}(k)$ of particles and antiparticles change through both channels of Table~\ref{tab:GMG2}. The pairwise channel creates or annihilates a
particle and an antiparticle together [Ref.~\cite{Xue2026b}, Sect.~4.2, Eq.~(4.4)]. The same-sign channel creates or annihilates two particles, or two antiparticles. The time-derivative coupling
shifts the energies to $\Omega\mp\mu$, which enters the detailed-balance factors. For each mode $k$, with $f_{\Phi,\Phi^*}\equiv f_{\Phi,\Phi^*}(k)$ and the per-mode rates \eqref{eq:golden} (Appendix~\ref{app:kinetic}),
\begin{equation}
\begin{aligned}
  \dot f_\Phi&=\Gamma_{M,k}^\uparrow(1+f_\Phi)(1+f_{\Phi^*})-\Gamma_{M,k}^\downarrow f_\Phi f_{\Phi^*}
   +2\Big[\Gamma_{C,k}^\uparrow e^{+2\mu/T_S}(1+f_\Phi)^2-\Gamma_{C,k}^\downarrow f_\Phi^2\Big],\\
  \dot f_{\Phi^*}&=\Gamma_{M,k}^\uparrow(1+f_\Phi)(1+f_{\Phi^*})-\Gamma_{M,k}^\downarrow f_\Phi f_{\Phi^*}
   +2\Big[\Gamma_{C,k}^\uparrow e^{-2\mu/T_S}(1+f_{\Phi^*})^2-\Gamma_{C,k}^\downarrow f_{\Phi^*}^2\Big],
\end{aligned}
\label{eq:kin_DQ2}
\end{equation}
where $T_S$ is the temperature of the spacetime vacuum, defined by detailed balance [Eq.~\eqref{eq:detbal}; KMS condition, Sect.~\ref{sec:kms}], and the factor $2$ counts the two quanta of a same-sign pair.
Three properties follow.
\begin{enumerate}[label=(\roman*),leftmargin=1.6em,itemsep=2pt]
\item \emph{Charge.} The pairwise terms are identical in both equations, so they conserve $f_\Phi-f_{\Phi^*}$; only the same-sign channel changes it.
\item \emph{Equilibrium.} Both channels vanish for $f_{\Phi,\Phi^*}=[e^{(\Omega\mp\mu)/T_S}-1]^{-1}$ (verified symbolically), so the equilibrium asymmetry per pair is
\begin{equation}
  \delta_{\rm eq}=\frac{f_\Phi-f_{\Phi^*}}{f_\Phi+f_{\Phi^*}}\simeq\frac{\mu}{T_S}\,(1+f^*)\simeq\frac{\mu}{T_S}\qquad(f^*\ll1,\ \mu\ll T_S),
  \label{eq:deltaeq}
\end{equation}
with $f^*$ the symmetric equilibrium occupation.
\item \emph{Relaxation.} Linearising about the equilibrium, the charge $f_\Phi-f_{\Phi^*}$ of mode $k$ relaxes towards $\delta_{\rm eq}(f_\Phi+f_{\Phi^*})$ at the rate $\Gamma_{C,k}=4\Gamma_{C,k}^\downarrow f_k^*$. Its average over
the modes is the macroscopic rate $\Gamma_C$ (Appendix~\ref{app:kinetic}).
\end{enumerate}
The bias requires a finite $T_S$, i.e.\ an asymmetric spectrum, $S_{\mathcal O}(+2\Omega)>S_{\mathcal O}(-2\Omega)$. The same-sign channel must therefore couple to the fluctuations of the spacetime vacuum, as
\eqref{eq:DeltaL} does.

\paragraph{Macroscopic form.} Integrating over the modes (Appendix~\ref{app:kinetic}), the charge density $n_Q=n_\Phi-n_{\Phi^*}$ obeys
$\dot n_Q+3Hn_Q=-\Gamma_C(n_Q-\delta_{\rm eq}n_M)$. With tracking, $d(a^3n_M)/dt=(3-2\epsilon)Ha^3n_M$, so the asymmetry per pair $\delta=n_Q/n_M$ satisfies
\begin{equation}
  \dot\delta=-(3-2\epsilon)H\,\delta-\Gamma_C\,(\delta-\delta_{\rm eq}),\qquad
  \delta\to\delta_{\rm ss}=\frac{\Gamma_C}{\Gamma_C+(3-2\epsilon)H}\,\delta_{\rm eq}.
  \label{eq:delta_eq_ode}
\end{equation}
The first term is the dilution of Eq.~\eqref{eq:dilution}, and the second is the new production. For $\Gamma_C\gg H$ the asymmetry tends to $\delta_{\rm eq}$,
\emph{whatever its initial value and whatever $N_0$} (R3). Each new generation of pairs is produced with the same bias, so tracking no longer dilutes the
asymmetry. Equation \eqref{eq:delta_eq_ode} adds to the closed set [Ref.~\cite{Xue2026b}, Eq.~(8.1)] without changing it, since $\delta\ll1$ leaves the pair energy density
unchanged (R5).

\paragraph{Charge equation along the cosmic history.}

With \eqref{eq:G2kappa}, the charge equation \eqref{eq:delta_eq_ode} becomes, in terms of $x=\ln a$,
\begin{equation}
  \frac{\dd\delta}{\dd x}=-(3-2\epsilon)\,\delta-\kappa_C\frac{\GaM}{H}\,\big(\delta-\delta_{\rm eq}\big).
  \label{eq:charge_grav}
\end{equation}
We solve it from the exit of today's scale, $N_0\simeq58.9$ e-folds before the end of inflation, through the $\mathcal P$- and $\mathcal M$-episodes, on the background of Refs.~\cite{XuePRD,Xue2023b,Xue2024}
(Table \ref{tab:baseline}, $r=0.022$, $g_Y^2=10^{-12}$). During inflation $\GaM/H=e^{\epsilon x}\le1$, with $x<0$. In the $\mathcal M$-episode $\GaM/H=1.2\times10^4$--$1.7\times10^7$. The channel
switches off by itself when the cosmological term is exhausted at $x_L=7.49$, because pair production stops: freeze-out needs no further assumption. In pre-inflation and inflation, $\GaM<H$, and their contributions are negligible.

\begin{figure}[t]
\centering
\includegraphics[width=\linewidth]{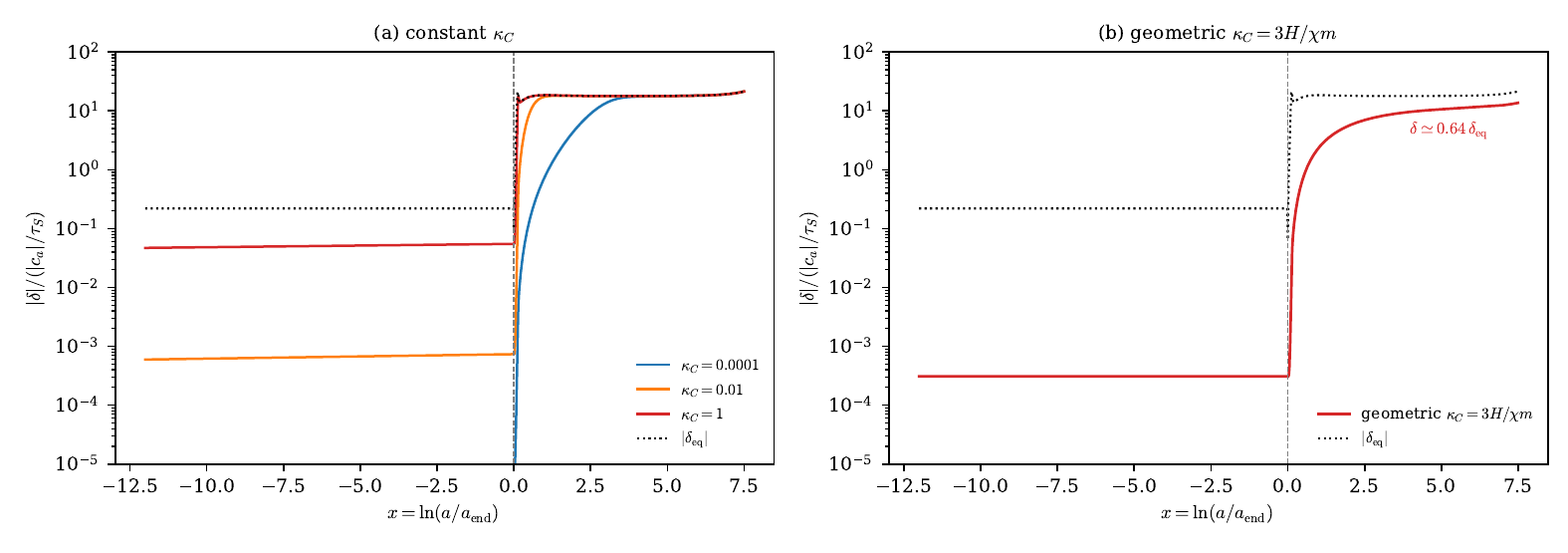}
\caption{Asymmetry per pair from Eq.~\eqref{eq:charge_grav} along the cosmic history of Refs.~\cite{XuePRD,Xue2023b,Xue2024} (Table \ref{tab:baseline}, $r=0.022$, $g_Y^2=10^{-12}$, $\tau_S=1$), against $x=\ln(a/a_{\rm end})$, for the scale-free time bias
$f=c_a\ln|R|$. The dashed vertical line marks the end of inflation, and the dotted line is $|\delta_{\rm eq}|$. (a) Constant $\kappa_C$: all values reach the same asymmetry at freeze-out,
$\delta\simeq-21.4\,c_a/\tau_S$. (b) Geometric $\kappa_C=3H/\chi m$, Eq.~\eqref{eq:k2geo}: since $\kappa_C\GaM/H\simeq0.34<1$ in the $\mathcal M$-episode, the asymmetry reaches $\delta\simeq0.64\,\delta_{\rm eq}$ at freeze-out.}
\label{fig:gravbias}
\end{figure}

\paragraph{Solution for constant $\kappa_C$.}

The asymmetry at freeze-out is independent of $\kappa_C$ over four decades [Fig.~\ref{fig:gravbias}(a)]:
\begin{center}\small
\begin{tabular}{@{}lccccc@{}}
\toprule
$\kappa_C$ & $10^{-4}$ & $10^{-3}$ & $10^{-2}$ & $10^{-1}$ & $1$\\
\midrule
$\delta/(c_a/\tau_S)$ at the end of inflation & $-2\times10^{-5}$ & $-1\times10^{-4}$ & $-7\times10^{-4}$ & $-7\times10^{-3}$ & $-0.055$\\
$\delta/(c_a/\tau_S)$ at freeze-out & $-21.4$ & $-21.4$ & $-21.4$ & $-21.4$ & $-21.4$\\
\bottomrule
\end{tabular}
\end{center}
Even for $\kappa_C=10^{-4}$ the channel is fast in the $\mathcal M$-episode, $\kappa_C\GaM/H\ge1.2$, so the asymmetry reaches the attractor whatever happened during inflation. It ends slightly above
$|\delta_{\rm eq}|=17.9\,c_a/\tau_S$ because $\epsilon$ rises towards $2$ just before freeze-out. With $n_B/s=0.91\,c_B(T_{\rm RH}/m_{\mathcal M})\,\delta$ (Sect.~\ref{sec:asymM}), the observed value requires
\begin{equation}
  \frac{|c_a|}{\tau_S}\simeq6.5\times10^{-5}\quad(c_B=1),\qquad1.9\times10^{-4}\quad(B-L),
  \label{eq:cb_req}
\end{equation}
for $T_{\rm RH}\simeq6\times10^{12}$\,GeV, and $|c_a|/\tau_S\simeq6.5\times10^{-6}$--$1.9\times10^{-3}$ over $T_{\rm RH}=6\times10^{11}$--$6\times10^{13}$\,GeV. Matter, rather than antimatter, requires $c_a<0$ in our
convention.

\paragraph{Solution for the geometric $\kappa_C$.}\label{sec:effL_geo}

With $\kappa_C$ from Eq.~\eqref{eq:k2geo}, the fraction of production events that straddle a horizon, Eq.~\eqref{eq:charge_grav} gives the evolution shown in Fig.~\ref{fig:gravbias}(b):
\begin{center}\small
\begin{tabular}{@{}lccccc@{}}
\toprule
$x=\ln(a/a_{\rm end})$ & $0$ & $2$ & $4$ & $6$ & $7.49$ (freeze-out)\\
\midrule
$\kappa_C\GaM/H$ & $0.004$ & $0.333$ & $0.339$ & $0.342$ & $0.364$\\
$\delta/(c_a/\tau_S)$ & $0.000$ & $-5.66$ & $-9.67$ & $-11.4$ & $-13.7$\\
$\delta_{\rm eq}/(c_a/\tau_S)$ & $-0.06$ & $-18.1$ & $-17.9$ & $-18.2$ & $-21.4$\\
\bottomrule
\end{tabular}
\end{center}
During inflation $\kappa_C\GaM/H\simeq0.004$, and no asymmetry builds up. In the $\mathcal P$-episode $\kappa_C\GaM/H$ rises to the plateau $3\epsilon/4\pi\simeq0.34$ and stays there through
the $\mathcal M$-episode, although $\kappa_C$ itself falls by four orders of magnitude, because $\kappa_C\propto H/m$ and $\GaM/H\propto m/H$. Since $\kappa_C\GaM/H<1$, the asymmetry approaches the attractor
only slowly and reaches $64\%$ of it by freeze-out [Fig.~\ref{fig:gravbias}(b)], rather than the full value obtained for $\kappa_C\GaM/H\gg1$.

 The observed $n_B/s$ then requires
\begin{equation}
  \frac{|c_a|}{\tau_S}\simeq1.0\times10^{-4},\qquad\text{i.e.}\qquad b_0\simeq1.3\times10^{-3}
  \qquad(c_B=1,\ T_{\rm RH}\simeq6\times10^{12}\,{\rm GeV}),
  \label{eq:b0_req}
\end{equation}
about $1.6$ times the value \eqref{eq:cb_req} for $\kappa_C\gtrsim10^{-4}$ constant. In the $\mathcal M$-episode, $\epsilon\simeq1.42$, so each horizon separation favours particles by about $2\times10^{-3}$.

\subsection{Discussion}

\begin{enumerate}[label=(\roman*),leftmargin=1.6em,itemsep=2pt]
\item \emph{A bias from the expansion.} The scale-free choice realises a bias set by the slow component $\dot H/H$ of the expansion. Its strength is a single dimensionless number,
$|c_a|/\tau_S\sim10^{-4}$, and for $\kappa_C\gtrsim10^{-4}$ the result does not depend on $\kappa_C$. The sign of the asymmetry, its homogeneity, the freeze-out and the comparison with observations are
discussed in Sect.~\ref{sec:asymM}.
\item \emph{Open points.} The temperature $T_S$ of the spacetime vacuum ($\tau_S$), the coefficient $c_a$ and the microscopic origin of $\kappa_C$ require a theory of the spacetime vacuum. The scale-free
choice involves $\ln|R|$, which is singular at $R=0$ in the radiation era. It is needed only while pair and charge production are active, $\epsilon<2$, and should be regarded as an effective
description of the bias during that period.
\end{enumerate}

\section{Asymmetry from charge-biased production in the \texorpdfstring{$\mathcal M$}{M}-episode}\label{sec:asymM}

Section~\ref{sec:transition} established the background from inflation to reheating. Section~\ref{sec:horizons} showed that horizon crossing, during inflation or at reheating, does not
produce the observed asymmetry. Section~\ref{sec:effL} constructed a Lagrangian for charge-biased pair production. Here, we analyse the asymmetry generated by the Lagrangian \eqref{eq:DeltaL} among
the real pairs of the $\mathcal M$-episode in reheating, and compare the result with observations.

\subsection{Setting and equations}

\paragraph{Ingredients.} We take the background of Sect.~\ref{sec:transition} (Table \ref{tab:baseline}, baseline $r=0.022$, $\chi=9.9\times10^{-4}$, $m_*=4.21\,\mpl$, $m_{\mathcal M}=37.9\,\mpl$, $g_Y^2=10^{-12}$) and add the terms
\eqref{eq:DeltaL}.
\begin{enumerate}[label=(\roman*),leftmargin=1.6em,itemsep=2pt]
\item \emph{Charge carriers.} Only the real pairs, $\rho^{\rm r}=\rho_M-\rhoMH$, carry charge; the polarisation $\rhoMH$ is neutral. The number density of charged particles is
$n^{\rm r}=n_\Phi+n_{\Phi^*}=\rho^{\rm r}/m_{\mathcal M}$. The densities $\rhoMH$ and $\rho^{\rm r}$ are computed with the collective mass $m_{\mathcal M}$, $m_{\mathcal M}^2=g_*M^2$, which includes the contributions of all $g_*$ degenerate states
(Sect.~\ref{sec:Leff}), so the number density follows with the same $m_{\mathcal M}$.
\item \emph{U(1)-violating rate.} The same-sign channel couples to the same fluctuations of the spacetime vacuum that produce the pairs, so its rate is proportional to the
pair-production rate, $\Gamma_C=\kappa_C\GaM$ with $\kappa_C\propto g_c^2$. It operates while the spacetime vacuum supplies energy ($\rho_\Lambda>0$). It switches off when the cosmological term is
exhausted at $x_\Lambda$, where pair production also stops. This is a natural freeze-out, requirement (R4) of Sect.~\ref{sec:newL}.
\item \emph{Bias.} The time-derivative coupling of \eqref{eq:DeltaL} gives particles and antiparticles the chemical potentials $\pm\mu$, $\mu=\dot f(R)$, Eq.~\eqref{eq:mu_R}. The equilibrium asymmetry
per particle is $\delta_{\rm eq}=\mu/T_S$, Eq.~\eqref{eq:deltaeq}. For the scale-free bias it is constant in the $\mathcal M$-episode, $\delta_{\rm eq}\simeq-17.9\,c_a/\tau_S$, and we treat it as a constant here.
\item \emph{Decays.} The Yukawa decays $\Phi\to\psi\bar N$ and $\Phi^*\to\bar\psi N$ transfer the charge of the decaying pairs to the fermions $\psi$, which carry U(1) charge $\pm1$ (Sect.~\ref{sec:yukawa}).
\end{enumerate}

\paragraph{Equations.} Let $N_\Phi=a^3n_\Phi$ be the comoving number of charged particles, $N_Q^\Phi$ and $N_Q^\psi$ the comoving charges carried by the pairs and by the decay products,
and $x=\ln(a/a_{\rm end})$. Then
\begin{equation}
  \frac{dN_Q^\Phi}{dx}=-\frac{\Gamma_C}{H}\big(N_Q^\Phi-\delta_{\rm eq}N_\Phi\big)-\frac{\Gde}{H}N_Q^\Phi,\qquad
  \frac{dN_Q^\psi}{dx}=\frac{\Gde}{H}N_Q^\Phi ,
  \label{eq:charge_M}
\end{equation}
with $\Gamma_C=0$ for $x>x_\Lambda$. The first term of the first equation is the biased relaxation of Sect.~\ref{sec:newL}. The dilution by neutral pairs from tracking enters through the growth of
$N_\Phi$. The decays move charge from the pairs to $\psi$ and conserve the total. The observable is the charge-to-entropy ratio after reheating,
\begin{equation}
  Y_Q=\frac{N_Q^\Phi+N_Q^\psi}{a^3s}\Big|_{\rm after\ reheating},\qquad s=\frac43\frac{\rho_R}{T},\qquad \rho_R=\frac{\pi^2}{30}g_{\rm eff}T^4,
  \label{eq:YQ}
\end{equation}
where $T$ is the temperature of the radiation after reheating, i.e.\ of the thermalised decay products, and $g_{\rm eff}$ its number of relativistic degrees of freedom. $Y_Q$ is
conserved afterwards. With a conversion factor $c_B$ from the U(1) charge to baryon number, $n_B/s=c_BY_Q$. We take $c_B=1$ if the charge is baryon number, and $c_B=28/79$ if it is
$B-L$, reprocessed by electroweak sphalerons.

\subsection{Evolution of the asymmetry and freeze-out}

\paragraph{Evolution of the asymmetry.}

Figure~\ref{fig:asym}(a) shows the asymmetry per real particle, $\delta=N_Q^\Phi/N_\Phi$. Its evolution has three stages.
\begin{itemize}[leftmargin=1.5em,itemsep=2pt]
\item \emph{$\mathcal P$-episode.} Real pairs appear, and $\delta$ rises towards the attractor at a rate set by $\Gamma_C/H=\kappa_C\GaM/H$.
\item \emph{$\mathcal M$-episode.} Since $\GaM/H\sim10^4$--$10^7$, $\delta=\delta_{\rm eq}$ to four digits for any $\kappa_C\ge10^{-4}$. Each new generation of real pairs is produced with the same bias, so tracking does not
dilute the asymmetry.
\item \emph{Freeze-out and transfer.} At $x_\Lambda=7.49$ the U(1)-violating channel switches off with $\delta=\delta_{\rm eq}$ exactly. The comoving charge is then conserved, and the decays transfer it to $\psi$ by the
time radiation dominates.
\end{itemize}
The result is independent of $\kappa_C$ over four decades:
\begin{equation}
  Y_Q=6.08\times10^{-8}\,\delta_{\rm eq}\qquad(\kappa_C=10^{-4},\ 10^{-2},\ 1),
  \label{eq:YQ_num}
\end{equation}
which can be written as
\begin{equation}
  Y_Q\simeq0.91\,\frac{T_{\rm RH}}{m_{\mathcal M}}\,\delta_{\rm eq}.
  \label{eq:YQ_an}
\end{equation}
The sudden-decay estimate $Y_Q=\tfrac34(T_{\rm RH}/m_{\mathcal M})\delta_{\rm eq}$ follows if all the energy of the pairs goes into radiation at $T_{\rm RH}$. It is smaller by a factor $1.2$, because part of the
radiation is produced earlier, when the Universe is hotter, and is subsequently diluted.

\paragraph{Freeze-out prescription.}

What happens after $x_\Lambda$ depends on whether the U(1)-violating channel and the bias persist.
\begin{center}\small
\begin{tabular}{@{}lcc@{}}
\toprule
after $x_\Lambda$ & $Y_Q/\delta_{\rm eq}$ & $\delta_{\rm eq}$ needed ($c_B=1$)\\
\midrule
$\Gamma_C=0$ (channel off with production) & $6.09\times10^{-8}$ & $1.43\times10^{-3}$\\
$\Gamma_C=\kappa_C\GaM$, $\mu$ persists & $6.16\times10^{-8}$ & $1.41\times10^{-3}$\\
$\Gamma_C=\kappa_C\GaM$, $\mu=0$ (washout of the pairs' charge) & $5.17\times10^{-9}$ & $1.68\times10^{-2}$\\
\bottomrule
\end{tabular}
\end{center}
The natural case, in which the channel stops with production, and the case in which the bias persists, agree to $1\%$. Washout of the pairs' charge, by a U(1)-violating channel that
outlives the bias, would cost a factor $12$. Only the charge already transferred to $\psi$ would survive. The result is robust as long as the bias lasts at least as long as the
U(1) violation.

\subsection{Required bias}

\paragraph{Dependence on the parameters.}

\begin{figure}[t]
\centering
\includegraphics[width=\linewidth]{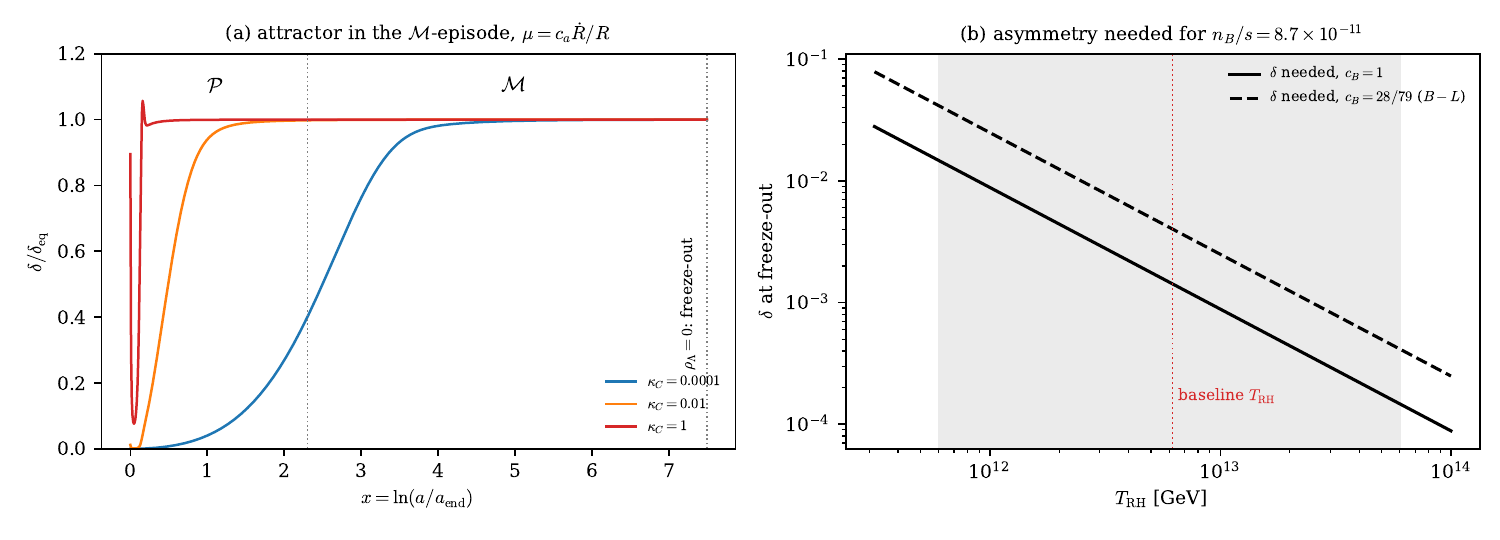}
\caption{(a) Asymmetry per pair relative to the attractor, $\delta/\delta_{\rm eq}$, from Eq.~\eqref{eq:charge_grav} with the scale-free bias, through the $\mathcal P$- and $\mathcal M$-episodes, for three values of $\kappa_C$.
All reach the attractor before freeze-out at $\rho_\Lambda=0$. (b) Asymmetry at freeze-out needed for $n_B/s=8.7\times10^{-11}$, against $T_{\rm RH}$, for $c_B=1$ and $c_B=28/79$; the shaded band is the range of
Refs.~\cite{XuePRD,Xue2023b,Xue2024}, and the dotted line its baseline.}
\label{fig:asym}
\end{figure}

Setting $c_BY_Q=n_B/s=8.7\times10^{-11}$, the value from $\Omega_bh^2=0.0224$, determines the bias. The reheating temperature controls it through Eq.~\eqref{eq:YQ_an}:
\begin{center}\small
\begin{tabular}{@{}lccccc@{}}
\toprule
$g_Y^2$ & $T_{\rm RH}$ [GeV] & $Y_Q/\delta_{\rm eq}$ & sudden decay & $\delta_{\rm eq}$ ($c_B=1$) & $\delta_{\rm eq}$ ($c_B=28/79$)\\
\midrule
$10^{-14}$ & $6.1\times10^{11}$ & $6.14\times10^{-9}$ & $5.1\times10^{-9}$ & $1.4\times10^{-2}$ & $4.0\times10^{-2}$\\
$10^{-12}$ & $6.2\times10^{12}$ & $6.09\times10^{-8}$ & $5.0\times10^{-8}$ & $1.4\times10^{-3}$ & $4.0\times10^{-3}$\\
$10^{-10}$ & $6.2\times10^{13}$ & $6.03\times10^{-7}$ & $5.0\times10^{-7}$ & $1.4\times10^{-4}$ & $4.1\times10^{-4}$\\
\bottomrule
\end{tabular}
\end{center}
The other parameters matter little. Over the allowed range of $\epsilon_{\mathcal M}$ ($1.05$--$1.42$) the required $\delta_{\rm eq}$ changes by less than $2\%$. Over the allowed range of $r$ it changes from
$1.3\times10^{-3}$ ($r=0.036$, $m_{\mathcal M}=29.6\,\mpl$) to $2.1\times10^{-3}$ ($r=0.005$, $m_{\mathcal M}=79.4\,\mpl$), following $m_{\mathcal M}/T_{\rm RH}$. The observed asymmetry thus requires a bias
\begin{equation}
  \delta_{\rm eq}=\frac{\mu}{T_S}\simeq1.1\,\frac{n_B/s}{c_B}\,\frac{m_{\mathcal M}}{T_{\rm RH}}\;\simeq\;1.4\times10^{-4}\text{--}4\times10^{-2}
  \label{eq:deltaeq_req}
\end{equation}
over $T_{\rm RH}=6\times10^{11}$--$6\times10^{13}$\,GeV ($g_Y^2=10^{-14}$--$10^{-10}$) and $c_B=28/79$--$1$ [Fig.~\ref{fig:asym}(b)]. A small bias, $\delta_{\rm eq}\ll1$, is sufficient everywhere in the parameter space of
Sect.~\ref{sec:transition}.

\paragraph{Required coefficient of the time bias.}

With the scale-free bias, the asymmetry at freeze-out is $\delta\simeq-21.4\,c_a/\tau_S$ for $\kappa_C\gtrsim10^{-4}$ and $-13.7\,c_a/\tau_S$ for the geometric $\kappa_C$ (Sect.~\ref{sec:effL}). Equation \eqref{eq:deltaeq_req}
then gives $|c_a|/\tau_S\simeq6.5\times10^{-5}$ and $1.0\times10^{-4}$ for the baseline (Table~\ref{tab:baseline}) ($c_B=1$), and $6.5\times10^{-6}$--$1.9\times10^{-3}$ over the whole parameter range. Only the value of $\delta_{\rm eq}$ near freeze-out matters, since the asymmetry tracks the attractor until $x_\Lambda$.

\subsection{Comparison with observations and conclusions}

\paragraph{Comparison with observations.}

\begin{enumerate}[label=(\roman*),leftmargin=1.6em,itemsep=2pt]
\item \emph{Baryon asymmetry.} The observed $n_B/s=8.7\times10^{-11}$, from $\Omega_bh^2=0.0224$ (CMB) and consistent with the deuterium abundance (BBN), is reproduced for $\delta_{\rm eq}$ given by
Eq.~\eqref{eq:deltaeq_req}. For the baseline (Table~\ref{tab:baseline}) ($T_{\rm RH}=6.2\times10^{12}$\,GeV) this is $\delta_{\rm eq}=1.4\times10^{-3}$ ($c_B=1$) or $4.0\times10^{-3}$ ($c_B=28/79$).
\item \emph{Sign and homogeneity.} The sign is that of $c_a\dot H$, fixed by the arrow of time, $\dot H<0$. It is the same throughout the observable Universe, and there are no matter--antimatter
domains, in agreement with the absence of annihilation signatures.
\item \emph{Baryon isocurvature.} The bias is built from the background geometry and has no field fluctuations of its own. It is modulated only by the adiabatic curvature perturbation,
so it adds no baryon isocurvature, consistently with the Planck bound $\beta_{\rm iso}<0.035$~\cite{Planck2018X}.
\item \emph{No washout after reheating.} The U(1)-violating operator acts only on $\Phi$, whose quanta are heavy ($M\gg T_{\rm RH}$), so its processes are Boltzmann-suppressed by $e^{-M/T}$, with $T$ the temperature of the plasma. Inverse
decays are suppressed in the same way, and the $\psi$ sector conserves the charge. If the charge is $B-L$, electroweak sphalerons convert it with $c_B=28/79$ and do not erase it.
\item \emph{Background and inflation unchanged.} $\delta\ll1$ leaves the pair energy density unchanged, so $n_s$, $r$, $N_*$ and $T_{\rm RH}$ of Sect.~\ref{sec:transition} are not affected.
\item \emph{CPT violation.} The expanding background breaks CPT, and the coupling $-\partial_\mu f(R)J^\mu$ turns this into a bias. Today $\mu\simeq-2c_a\epsilon H_0$ is of order $10^{-37}$\,GeV, and the
U(1)-violating channel has long frozen out, so there is no conflict with laboratory bounds.
\end{enumerate}

\paragraph{Washout.}\label{sec:washout} After freeze-out the same-sign channel is off, since pair production has stopped. The remaining processes conserve $Q_X$. Inverse decays $\psi\bar N\to\Phi$ are suppressed by $e^{-M/T}$, with $T\le T_{\rm RH}$ the temperature of the plasma after reheating, for $T\ll M$.
With $T_{\rm RH}\simeq6\times10^{11}$--$6\times10^{13}$\,GeV for $g_Y^2=10^{-14}$--$10^{-10}$, the mass scale $M$ of the decaying pairs must exceed $T_{\rm RH}$ by a factor of order ten or more. Electroweak sphalerons then act as in Eq.~\eqref{eq:sphaleron} (Sect.~\ref{sec:baryo}).

\paragraph{Consistency of the rates.} In the cosmic history of Refs.~\cite{XuePRD,Xue2023b,Xue2024} the rates are ordered as required. During inflation $\GaM<H$, and the polarisation follows the expansion
adiabatically. In the $\mathcal M$-episode $\GaM/H\sim10^4$--$10^7$, the pairs are in detailed balance, and the same-sign channel reaches the attractor. The decays become effective, $\Gde>H$, only in
the $\mathcal R$-episode, after the cosmological term is exhausted and the asymmetry has frozen. The decay rate $\Gde=g_Y^2m_{\mathcal M}$ with $g_Y^2=10^{-12}$ corresponds to a Yukawa coupling $y\simeq5\times10^{-6}g_*^{1/4}$
(Sect.~\ref{sec:rates}).

\paragraph{Conclusions.}

\begin{enumerate}[label=(\roman*),leftmargin=1.6em,itemsep=2pt]
\item \emph{The $\mathcal M$-episode is the natural site of the asymmetry.} Real pairs dominate the energy density, and $\GaM/H\sim10^4$--$10^7$ keeps the biased U(1)-violating channel in
equilibrium. The asymmetry per particle then sits at the attractor $\delta_{\rm eq}=\mu/T_S$, independent of the coupling $\kappa_C$ and of the initial state.
\item \emph{Freeze-out is automatic.} When the cosmological term is exhausted, pair production and the U(1)-violating channel stop together. The decays then carry the frozen charge into the
radiation, giving $Y_Q\simeq0.91\,(T_{\rm RH}/m_{\mathcal M})\,\delta_{\rm eq}$.
\item \emph{The observed asymmetry needs a small bias.} $n_B/s=8.7\times10^{-11}$ requires $\delta_{\rm eq}\simeq1.4\times10^{-3}$ ($c_B=1$) or $4\times10^{-3}$ ($B-L$) for the baseline (Table~\ref{tab:baseline}), and $1.4\times10^{-4}$--$4\times10^{-2}$ over the
whole parameter space. For the scale-free bias this means $|c_a|/\tau_S\sim10^{-4}$.
\item \emph{The scenario agrees with all observations considered:} $n_B/s$, the absence of antimatter domains, the isocurvature bound, the absence of washout, the unchanged
inflationary and reheating observables, and the absence of present-day CPT violation.
\end{enumerate}

\section{Transfer to baryons, leptons and dark matter}\label{sec:transfer}\label{sec:baryolepto}

Section~\ref{sec:asymM} gave the charge asymmetry of the pairs at freeze-out, $Y_Q\simeq0.91\,(T_{\rm RH}/m_{\mathcal M})\,\delta$, with a sign fixed everywhere by the bias $\mu$. In the $\mathcal R$-episode the pairs decay, and
the asymmetry passes to light particles. This section derives the resulting baryon, lepton and darkon asymmetries from the quantum numbers of the decay products.

\subsection{Decays and quantum numbers}\label{sec:quantumnumbers}

\paragraph{The Yukawa coupling.} The decays are mediated by the U(1)-invariant Yukawa term of the Lagrangian \eqref{eq:Leff}, $-y(\Phi\bar\psi N+{\rm h.c.})$, with $\psi$ of U(1) charge $+1$ and $N$ neutral. It gives
\begin{equation}
  \Phi\to\psi+\bar N,\qquad \Phi^*\to\bar\psi+N .
  \label{eq:decayBL}
\end{equation}
The pair annihilations $\Phi\Phi^*\to\psi\bar\psi,\,N\bar N,\dots$ carry no net number. Every additive quantum number $X$ conserved by the Yukawa term satisfies
\begin{equation}
  X_\Phi=X_\psi-X_N,\qquad X\in\{B,\,L,\,D\},
  \label{eq:qnumbers}
\end{equation}
where $B$, $L$ and $D$ are the baryon, lepton and darkon numbers. The U(1) charge of $\Phi$ is thus carried by $\psi$, and its $B$, $L$ and $D$ numbers are those of the pair $\psi\bar N$. Table~\ref{tab:channels} lists
the channels.

\begin{table}[t]
\centering\small
\begin{tabular}{@{}llcccl@{}}
\toprule
channel & decay $\Phi\to\psi+\bar N$ & $(B,L,D)_\Phi$ & $q_{B-L}$ & $q_D$ & result\\
\midrule
$V_\ell$: lepton & $\Phi\to\ell+\bar N^0_V$ & $(0,1,0)$ & $-1$ & $0$ & leptogenesis\\
$V_{q\bar\ell}$: quark, antilepton & $\Phi\to q+\bar\ell$ & $(\tfrac13,-1,0)$ & $\tfrac43$ & $0$ & baryogenesis\\
$V_{q\ell}$: quark, lepton & $\Phi\to q+\ell$ & $(\tfrac13,1,0)$ & $-\tfrac23$ & $0$ & baryogenesis\\
$V_0$: $B-L$ conserving & $B_\psi-B_N=L_\psi-L_N$ & $B_\Phi=L_\Phi$ & $0$ & $0$ & washed out\\
$D$: darkon & $\Phi\to\psi_D+\bar N^0_D$ & $(0,0,1)$ & $0$ & $1$ & darkogenesis\\
\bottomrule
\end{tabular}
\caption{Decay channels $\Phi\to\psi+\bar N$ of the Yukawa term of \eqref{eq:Leff}. Here $\ell$ is a lepton, $q$ a quark, $\psi_D$ a darkon (a dark fermion with $D=1$), and $N^0_V$, $N^0_D$ singlets. The quantum numbers of $\Phi$
follow from Eq.~\eqref{eq:qnumbers}, and $q_{B-L}$ and $q_D$ are the $B-L$ and $D$ numbers produced per decay of $\Phi$, Eq.~\eqref{eq:qdef}. The conjugate $\Phi^*$ produces the opposite. Electroweak
sphalerons preserve $B-L$ and erase $B+L$ (Sect.~\ref{sec:baryo}).}
\label{tab:channels}
\end{table}

\paragraph{Numbers produced per decay.} We define the $B-L$ and $D$ numbers produced by one decay of $\Phi$,
\begin{equation}
  q_{B-L}\equiv(B-L)_\Phi=(B-L)_\psi-(B-L)_N,\qquad q_D\equiv D_\Phi=D_\psi-D_N ,
  \label{eq:qdef}
\end{equation}
given for each channel in Table~\ref{tab:channels}. A visible channel produces $q_{B-L}\ne0$ and $q_D=0$; the dark channel produces $q_D\ne0$ and $q_{B-L}=0$.

\paragraph{Several channels.} Suppose $\Phi$ decays both into a visible channel V, with branching ratio ${\rm Br}_V$, and into the dark channel D, with ${\rm Br}_D$. No single assignment of $B-L$ and $D$ to
$\Phi$ can then satisfy Eq.~\eqref{eq:qnumbers} for both channels, since the products of V have $D=0$ and those of D have $B-L=0$. The quantity conserved by the Lagrangian is the U(1) charge of
$\Phi$, equal to $+1$ for every $\Phi$. Each decay transfers this unit of charge to its products, as $B-L=q_{B-L}$ in channel V or as $D=q_D$ in channel D. The products therefore carry the
generalised charge
\begin{equation}
  Q_X=\frac{B-L}{q_{B-L}}+\frac{D}{q_D},
  \label{eq:QX}
\end{equation}
which equals $+1$ for the products of either channel, the charge of the $\Phi$ that decayed. $Q_X$ is the U(1) charge of \eqref{eq:Leff} extended to the decay products. It is conserved by the
pairwise coupling, the Yukawa couplings and the decays, and violated only by the same-sign channel $g_c\Phi^2$, by two units with the sign fixed by the bias. $B-L$ and $D$ are conserved
within each channel, but an excess of $\Phi$ is shared between them according to the branching ratios. In a single channel, $Q_X$ reduces to $(B-L)/q_{B-L}$ or to $D/q_D$.

\paragraph{From the pair asymmetry to the decay products.} After the decays, the excess $Y_Q$ of $\Phi$ over $\Phi^*$ per entropy becomes
\begin{equation}
  \frac{n_{B-L}}{s}=q_{B-L}\,{\rm Br}_V\,Y_Q,\qquad \frac{n_D}{s}=q_D\,{\rm Br}_D\,Y_Q .
  \label{eq:nBnL}
\end{equation}

\subsection{Baryogenesis and leptogenesis}\label{sec:baryo}

\paragraph{Electroweak sphalerons.} Between about $10^{12}$\,GeV and the electroweak scale, sphaleron processes violate $B+L$ and conserve $B-L$~\cite{Kuzmin1985}. In chemical equilibrium they leave, for the
Standard Model with one Higgs doublet~\cite{HarveyTurner1990},
\begin{equation}
  \frac{n_B}{s}=\frac{28}{79}\,\frac{n_{B-L}}{s},\qquad \frac{n_L}{s}=-\frac{51}{79}\,\frac{n_{B-L}}{s}.
  \label{eq:sphaleron}
\end{equation}
The decays occur at $T_{\rm RH}\simeq6\times10^{12}$\,GeV (Table~\ref{tab:baseline}), before the sphaleron era, so Eq.~\eqref{eq:sphaleron} acts on the asymmetry they produce.
\begin{itemize}[leftmargin=1.5em,itemsep=2pt]
\item \emph{$B-L$-conserving decays} ($V_0$): $n_{B-L}=0$, and the sphalerons erase the asymmetry. This case is not viable unless the asymmetry is stored in a sector the sphalerons do not reach.
\item \emph{$B-L$-violating decays} ($V_\ell$, $V_{q\bar\ell}$, $V_{q\ell}$): $n_B/s=\tfrac{28}{79}\,q_{B-L}{\rm Br}_VY_Q$, the case $c_B=\tfrac{28}{79}|q_{B-L}|{\rm Br}_V$ of
Sect.~\ref{sec:asymM}.
\end{itemize}

\paragraph{Leptogenesis.} Channel $V_\ell$ decays into leptons only, in analogy with the decay of heavy neutrinos in standard leptogenesis~\cite{FukugitaYanagida1986,Buchmuller2005,Davidson2008}. The lepton
asymmetry $n_L/s=-n_{B-L}/s={\rm Br}_VY_Q$ is partly converted into baryons by the sphalerons, $n_B/s=-\tfrac{28}{79}{\rm Br}_VY_Q$, so matter requires $Y_Q<0$, i.e.\ an excess of $\Phi^*$. The origin of the
asymmetry is the difference from standard leptogenesis. Here the decays are CP-symmetric, and the asymmetry is that of the decaying particles themselves, produced by the biased
same-sign channel. The observed $n_B/s=8.7\times10^{-11}$ requires $|\delta|\simeq4\times10^{-3}/{\rm Br}_V$ at the baseline (Table~\ref{tab:baseline}), i.e.\ $|c_a|/\tau_S\simeq1.9\times10^{-4}/{\rm Br}_V$ for the scale-free bias with $\kappa_C\gtrsim10^{-4}$.

\paragraph{Charged and neutral components.} At $T_{\rm RH}$ all light states are relativistic. Weak interactions keep them in $\beta$-equilibrium: processes such as $u\leftrightarrow d+e^++\nu_e$ relate the chemical
potentials of the charged and neutral states, $\mu_{B_c}-\mu_{B_0}=\mu_{L_c}-\mu_{L_0}$. Electric neutrality, in addition, requires the charged baryon and lepton asymmetries to balance, $n_{B_c}\simeq n_{L_c}$. The baryon asymmetry is shared between charged and neutral baryons, which differ in mass by much less than the temperature $T_{\rm RH}$, so
$n_{B_c}\simeq n_B/2$. The neutrino asymmetry is then of the order of $n_B/s\sim10^{-10}$, far below the nucleosynthesis bound on the neutrino chemical potentials, $|\xi_\nu|\lesssim0.1$, i.e.\
$|n_{\nu}/s|\lesssim10^{-2}$.

\subsection{Darkogenesis}\label{sec:darko}

\paragraph{Visible and dark Yukawa couplings.} Let $\Phi$ couple to the visible and dark channels with two Yukawa couplings,
\begin{equation}
  \mathcal L_Y=-y_V\,\Phi\,\bar\psi_VN^0_V-y_D\,\Phi\,\bar\psi_DN^0_D+{\rm h.c.},
  \label{eq:LYVD}
\end{equation}
where $\psi_V$ is a visible fermion, $\psi_D$ a darkon, and $N^0_V$, $N^0_D$ are singlets (Table~\ref{tab:channels}). For relativistic final states the partial widths are approximately $y_{V,D}^2M/8\pi$. The branching ratios are therefore
\begin{equation}
  {\rm Br}_{V,D}=\frac{y_{V,D}^2}{y_V^2+y_D^2},\qquad \frac{{\rm Br}_D}{{\rm Br}_V}=\frac{y_D^2}{y_V^2},
  \label{eq:BrVD}
\end{equation}
and the total decay rate $\Gde=(y_V^2+y_D^2)M/8\pi$ defines $g_Y^2=(y_V^2+y_D^2)M/8\pi m_{\mathcal M}$ in Eq.~\eqref{eq:rates}.

\paragraph{Darkon asymmetry.} Channel D produces the darkon asymmetry $n_D/s=q_D{\rm Br}_DY_Q$, Eq.~\eqref{eq:nBnL}. It has the same sign as the visible asymmetry, since both come from the same bias, and it is
not processed by the sphalerons. This is asymmetric dark matter~\cite{Kaplan2009,Zurek2014}, generated by the biased production of the pairs.

\paragraph{The symmetric component.} Darkons and antidarkons are also produced symmetrically, by the annihilation $\Phi\Phi^*\to\psi_D\bar\psi_D$ and by scatterings in the plasma. This symmetric part
has the same origin as a thermal relic. If darkons annihilate into lighter states, either dark radiation or Standard Model particles through a mediator, with a cross-section much larger
than the thermal value, $\langle\sigma v\rangle\gg3\times10^{-26}\,{\rm cm^3\,s^{-1}}$, the symmetric part is depleted far below the observed density. Only the excess of darkons over antidarkons, which
annihilation cannot remove, then survives. The dark-matter density is then set by the asymmetry, $\rho_{\rm DM}=\langle m_D\rangle\,n_D$, where $\langle m_D\rangle$ is the average mass of the darkons if
there are several species.

\paragraph{Darkon mass.} The observed ratio $\Omega_{\rm DM}/\Omega_B=5.36$~\cite{Planck2018VI}, with $n_B/s$ from Eq.~\eqref{eq:sphaleron} and the branching ratios \eqref{eq:BrVD}, fixes the average darkon mass, with $m_p$ the proton mass:
\begin{equation}
  \langle m_D\rangle=5.36\,m_p\,\frac{n_B}{n_D}=5.36\,m_p\,\frac{28}{79}\,\frac{|q_{B-L}|\,{\rm Br}_V}{q_D\,{\rm Br}_D}\simeq1.8\,{\rm GeV}\times\frac{|q_{B-L}|}{q_D}\,\frac{y_V^2}{y_D^2}.
  \label{eq:mD}
\end{equation}
The two Yukawa couplings enter in two ways. Their sum sets the decay rate, and hence $T_{\rm RH}$ and $Y_Q$. Their ratio sets the darkon mass, because $Y_Q$ cancels in $n_B/n_D$. For equal couplings and unit
charges, $\langle m_D\rangle\simeq1.8$\,GeV, the typical scale of asymmetric dark matter. A heavier darkon requires a weaker dark Yukawa coupling, $y_D<y_V$. For example, $\langle m_D\rangle\simeq100$\,GeV needs
$y_D/y_V\simeq0.13$. A lighter one requires $y_D>y_V$.

\section{Magnetogenesis}\label{sec:magneto}

The decays \eqref{eq:decayBL} of the asymmetric pairs produce charged baryons and leptons. Their charges balance, but their velocities need not. The resulting electric current generates a
magnetic field at reheating.

\subsection{Electric current and Maxwell's equations}

\paragraph{Electric current from the decays.} In the visible channels of Table~\ref{tab:channels}, each decay of an excess $\Phi$ produces a charged baryonic state and a charged leptonic state of opposite
charges, with $n_{B_c}\simeq n_{L_c}$. The electric current density is
\begin{equation}
  \JJ_{\rm em}=e\,\big(n_{B_c}\mathbf v_B-n_{L_c}\mathbf v_L\big)=e\,n_{B_c}\,\Delta\mathbf v,\qquad \Delta\mathbf v=\mathbf v_B-\mathbf v_L ,
  \label{eq:jR}
\end{equation}
where $\mathbf v_B$ and $\mathbf v_L$ are the mean velocities of the two charged populations. In the rest frame of a decaying $\Phi$ the two products have equal and opposite momenta but different masses,
hence different speeds, so $\Delta\mathbf v\ne0$ for each decay. Averaged over random decay directions the current vanishes locally. A net current over a region requires a net flow of the decay
products through it, which the generation of the asymmetry sets up.

\paragraph{Maxwell's equations.} In the reheating plasma, at temperature $T\simeq T_{\rm RH}$, the electric conductivity is high, $\sigma\simeq T_{\rm RH}/(\alpha\ln\alpha^{-1})\sim10\,T_{\rm RH}$. The displacement current and
the electric field are therefore negligible on the Hubble scale, and Ampère's law reads $\nabla\times\mathbf B=4\pi\JJ_{\rm em}$. The magnetic diffusion time over the Hubble scale, $4\pi\sigma H_{\rm RH}^{-2}$,
exceeds the Hubble time by $4\pi\sigma/H_{\rm RH}\sim10^7$, so the field is frozen into the plasma. From then on, flux conservation gives $B\propto a^{-2}$.

\paragraph{The relative velocity.} The current depends on $|\Delta v|$, which lies between two limits.
\begin{itemize}[leftmargin=1.5em,itemsep=2pt]
\item \emph{Upper limit,} $|\Delta v|=1$: relativistic decay products moving in opposite directions.
\item \emph{Lower limit,} set by the flow that the asymmetry requires. In a Hubble sphere of radius $R=H^{-1}$, the decays of the excess $\Phi$ create charged products at the rate $\Gde\,\delta\,n$ per unit
volume, with $n$ the number density of the decaying pairs. A charge-separating flow must carry this excess through the surface of the sphere. Gauss's theorem for
$\nabla\cdot(n\,\Delta\mathbf v)=\Gde\,\delta\,n$ gives
\begin{equation}
  n\,|\Delta v|\,4\pi R^2=\Gde\,\delta\,n\,\frac{4\pi}{3}R^3\quad\Longrightarrow\quad |\Delta v|_{\min}=\frac{\Gde}{3H}\,\delta .
  \label{eq:dvmin}
\end{equation}
At reheating the decays complete when $t\simeq(\Gde)^{-1}$, and in the radiation era $H=1/2t$, so $\Gde\simeq2H_{\rm RH}$ and $|\Delta v|_{\min}\simeq\tfrac23\delta$. With $\delta\simeq1.4\times10^{-3}$ (Sect.~\ref{sec:asymM}),
$|\Delta v|_{\min}\simeq9\times10^{-4}$.
\end{itemize}

\subsection{The field at reheating and today}

\paragraph{At reheating.} Ampère's law integrated over a disc of radius $R=H_{\rm RH}^{-1}$ bounded by a Hubble sphere, with a homogeneous current, gives
\begin{equation}
  2\pi R\,B_R\simeq4\pi\,j_{\rm em}\,\pi R^2\quad\Longrightarrow\quad B_R\simeq\frac{2\pi e\,n_{B_c}|\Delta v|}{H_{\rm RH}},\qquad
  \frac{B_R}{B_c}\simeq\frac{2\pi\alpha}{m_e^2H_{\rm RH}}\,|\Delta v|\,n_{B_c},
  \label{eq:BR}
\end{equation}
in Gaussian units, with $e^2=\alpha$ and the critical field $B_c=m_e^2/e\simeq4.4\times10^{13}$\,G. For the baseline (Table~\ref{tab:baseline}), $T_{\rm RH}=6.2\times10^{12}$\,GeV, $H_{\rm RH}=5.3\times10^7$\,GeV and $g_{\rm eff}=106.75$.
The entropy density is then $s_R=(2\pi^2/45)g_{\rm eff}T_{\rm RH}^3=1.1\times10^{40}$\,GeV$^3$, and $n_{B_c}\simeq\tfrac12(n_B/s)s_R\simeq4.9\times10^{29}$\,GeV$^3$.

\paragraph{Today.} Entropy conservation gives $a_R/a_0=(g_{s,0}/g_{\rm eff})^{1/3}T_0/T_{\rm RH}\simeq1.3\times10^{-26}$, with $g_{s,0}=3.91$, and flux conservation $B_0=(a_R/a_0)^2B_R$:
\begin{center}\small
\begin{tabular}{@{}lccc@{}}
\toprule
 & $|\Delta v|$ & $B_R$ [G] & $B_0$ [G]\\
\midrule
upper limit & $1$ & $7\times10^{40}$ & $1\times10^{-11}$\\
lower limit, Eq.~\eqref{eq:dvmin} & $\tfrac23\delta\simeq9\times10^{-4}$ & $7\times10^{37}$ & $1\times10^{-14}$\\
\bottomrule
\end{tabular}
\end{center}
The range $10^{-14}\lesssim B_0\lesssim10^{-11}$\,G lies between the CMB upper bound, $B\lesssim10^{-9}$\,G on Mpc scales~\cite{Planck2015XIX}, and the lower bound from blazar observations,
$B\gtrsim10^{-17}$\,G~\cite{NeronovVovk2010}.

\paragraph{Coherence length and evolution.} The field is coherent over a Hubble volume at reheating. Redshifted to today, that length is $H_{\rm RH}^{-1}(a_0/a_R)\simeq3\times10^2$\,m, i.e.\ $10^{-20}$\,Mpc. The
bounds above refer to coherence lengths of order a Mpc. The comparison is therefore meaningful only after the magnetohydrodynamic evolution from reheating to today. That evolution
damps the field on small scales and, for a helical field, transfers power to larger scales by an inverse cascade~\cite{Grasso2001,BanerjeeJedamzik2004,DurrerNeronov2013,Subramanian2016}. In the
present framework the sign of the asymmetry is uniform, but the direction of the current varies from one Hubble volume to the next. The estimate above is therefore the amplitude at the
coherence length of reheating, and an upper bound on the field at larger scales. Whether a field of the observed strength survives on Mpc scales requires a study of its MHD evolution,
which we leave to future work.

\section{Conclusions and remarks}\label{sec:conclusions}

We have studied whether the separation of pair members by horizons, continuous from pre-inflation to reheating, can generate the observed particle--antiparticle asymmetry, using the
pair production and the cosmic history of Refs.~\cite{Xue2026b,XuePRD,Xue2023b,Xue2024}. We answer the two questions raised in the introduction.

\paragraph{(i) Charge separation at horizons.}
\begin{itemize}[leftmargin=1.5em,itemsep=2pt]
\item \emph{A single act of separation is not enough.} The charge of a patch at the exit of its scale has a random sign and is diluted by $10^{-50.5}$ by the subsequent tracking (Sect.~\ref{sec:horizons}).
\item \emph{If horizons only transport charge,} the charge of today's Hubble volume changes only through its boundary, by local charge conservation. The result is a surface charge, at most
$10^{51}$--$10^{52}$ even with a coherent sign, against the observed $2.6\times10^{78}$ (Sect.~\ref{sec:singlepatch}).
\item \emph{If horizons absorb the charge they separate,} as expected since gravity does not respect global symmetries, and if real pairs and charge are produced continuously, the
separations at all horizons inside today's Hubble volume act as a volume violation of the U(1) charge. Pairs straddle a horizon with probability $\kappa_C\simeq3H/\chi m$, so that $\kappa_C\GaM/H=3\epsilon/4\pi$ (Sect.~\ref{sec:manyhorizons}).
\item \emph{The sign} requires, besides the time arrow of the evolving horizon, a C- and CP-odd coupling (Sect.~\ref{sec:horizonbias}).
\end{itemize}

\paragraph{(ii) Effective Lagrangian and observations.} The U(1)-violating term $g_c\,\delta\Ocal_S\Phi^2$, with $\Gamma_C=\kappa_C\GaM$, and the time bias $-\partial_\mu f(R)\,J^\mu$ describe this
picture with no new field (Sect.~\ref{sec:effL}).
\begin{itemize}[leftmargin=1.5em,itemsep=2pt]
\item \emph{Bias from $\dot H/H$.} For $f=c_a\ln|R|$, $\mu\simeq2c_a\dot H/H$.
\item \emph{When the asymmetry forms.} The asymmetry per pair reaches the attractor $\delta_{\rm eq}=\mu/T_S$ in the pair-dominated episode and freezes when the cosmological term is exhausted, with no
assumed freeze-out. Pre-inflation and inflation do not contribute enough for the observed size, since $\Gamma_C<H$ there.
\item \emph{Required bias.} The observed $n_B/s=8.7\times10^{-11}$ fixes the asymmetry per pair at freeze-out, $\delta\simeq1.4\times10^{-3}$ for $c_B=1$ and the baseline $T_{\rm RH}\simeq6\times10^{12}$\,GeV
(Table~\ref{tab:baseline}). For the scale-free time bias this requires $|c_a|/\tau_S\simeq6.5\times10^{-5}$ if the charge channel is fast in the $\mathcal M$-episode ($\kappa_C\gtrsim10^{-4}$, independent of $\kappa_C$), and
$|c_a|/\tau_S\simeq1.0\times10^{-4}$ for the geometric $\kappa_C$, i.e.\ a bias of about $2\times10^{-3}$ per horizon separation. Over the allowed range, $T_{\rm RH}=6\times10^{11}$--$6\times10^{13}$\,GeV and $c_B=28/79$--$1$,
$|c_a|/\tau_S$ lies between $6.5\times10^{-6}$ and $1.9\times10^{-3}$.
\item \emph{Observations.} The scenario reproduces $n_B/s$, gives a sign fixed by the arrow of time with no antimatter domains, and adds no isocurvature. It has no washout after reheating,
leaves the inflationary and reheating observables of Refs.~\cite{XuePRD,Xue2023b,Xue2024} unchanged, and leaves no CPT violation today (Sect.~\ref{sec:asymM}).
\end{itemize}

\paragraph{(iii) Baryogenesis, leptogenesis, darkogenesis and magnetogenesis} (Sects.~\ref{sec:baryolepto}--\ref{sec:magneto}).
\begin{itemize}[leftmargin=1.5em,itemsep=2pt]
\item \emph{Baryogenesis and leptogenesis.} The Yukawa decays $\Phi\to\psi\bar N$ transfer the asymmetry of the pairs to baryons and leptons with a uniform sign, with the quantum numbers fixed by
$X_\Phi=X_\psi-X_N$. Electroweak sphalerons preserve it only if the decays carry $B-L\ne0$, giving $n_B/s=\tfrac{28}{79}q_{B-L}{\rm Br}_VY_Q$. A decay into leptons only realises leptogenesis, with the asymmetry from biased production instead of CP-violating decays.
\item \emph{Darkogenesis.} Decays into a dark sector give asymmetric dark matter with the same sign. $\Omega_{\rm DM}/\Omega_B=5.36$ then fixes $m_D\simeq1.8\,{\rm GeV}\times|q_{B-L}|{\rm Br}_V/(q_D{\rm Br}_D)$.
\item \emph{Magnetogenesis.} The charged decay products carry a current. At the coherence length of reheating, $\sim3\times10^2$\,m today, the field is $10^{-14}\lesssim B_0\lesssim10^{-11}$\,G. Comparison with the
Mpc-scale bounds requires its MHD evolution.
\end{itemize}

\paragraph{Remarks.}
\begin{enumerate}[label=(\roman*),leftmargin=1.6em,itemsep=2pt]
\item \emph{Horizons as charge sinks.} The absorption of charge by horizons is the key physical assumption. It is motivated by the absence of exact global symmetries in
gravity~\cite{Kallosh1995,BanksSeiberg2011}, but its derivation requires a microscopic theory of the spacetime vacuum.
\item \emph{The spacetime vacuum.} The temperature $T_S$ ($\tau_S$), the coefficient $c_a$, the coefficients of $\GaM$ and $\Gamma_C$, and the U(1)-violating spectral function also require such a theory.
\item \emph{Epoch-dependent mass function.} The increase of $m$ at the end of inflation, which sets $T_{\rm RH}/m_{\mathcal M}$ and hence the required bias, is fixed by observations~\cite{XuePRD,Xue2023b,Xue2024}. Its
microscopic mechanism requires the mass scale $M$ and the degeneracy $g_*$ of the pair states.
\item \emph{Late Universe.} The cosmological term is exhausted at reheating; the present dark energy then requires its regeneration in the late Universe~\cite{Xue2024}.
\end{enumerate}

\appendix

\section{Derivations for Section~\ref{sec:horizons}}\label{app:boundary}

\paragraph{Pairs straddling the boundary of today's Hubble volume.} After $N$ e-folds following the exit of today's scale, the comoving Hubble radius is $r_H=R_0e^{-(1-\epsilon)N}$. The boundary
$\partial V_0$, of area $4\pi R_0^2$, is covered by $4\pi R_0^2/\pi r_H^2=4(R_0/r_H)^2$ Hubble discs, each producing $\eta_r$ straddling pairs per e-fold. Summing,
\begin{equation}
  N_\partial=\int_0^{N_0}4\eta_r\,e^{2(1-\epsilon)N}\,\dd N=\frac{4\eta_r}{2(1-\epsilon)}\Big[e^{2(1-\epsilon)N_0}-1\Big],
  \label{eq:Nb_app}
\end{equation}
which gives Eq.~\eqref{eq:Nboundary}. With random signs the charge is $|Q_0|\simeq N_\partial^{1/2}$, and with a coherent sign $|Q_0|\le N_\partial$. After inflation $r_H$ grows, and the boundary adds less.

\paragraph{The required charge.} With $s_0=2891\,{\rm cm}^{-3}$ and $V_0=\tfrac{4\pi}{3}(c/H_0)^3=1.05\times10^{85}\,{\rm cm}^3$ ($c/H_0=4.4$\,Gpc), $s_0V_0=3.0\times10^{88}$, and $Q_{\rm req}=(n_B/s)\,s_0V_0=2.6\times10^{78}$ for
$n_B/s=8.7\times10^{-11}$.

\paragraph{The geometric $\kappa_C$.} In a sphere of radius $r=H^{-1}$, the fraction of the volume within $\lambda_m\ll r$ of the surface is $1-(1-\lambda_m/r)^3\simeq3\lambda_mH$. Pairs produced there straddle the
horizon, so $\kappa_C\simeq3\lambda_mH=3H/\chi m$. With $\GaM=\chi m\epsilon/4\pi$, $\kappa_C\GaM/H=3\epsilon/4\pi$, independent of $m$ and $H$.

\paragraph{Numerical solution.} Equation \eqref{eq:charge_grav} is integrated from $x=-N_0$ to $x_\Lambda$ with LSODA (relative tolerance $10^{-8}$, maximal step $0.05$ e-fold). For $x<0$ we use
$\epsilon=0.0175$, $\GaM/H=e^{\epsilon x}$ and $m=m_*$. For $x\ge0$, $\epsilon(x)$, $\GaM/H(x)$ and $m(x)$ are taken from the closed set of Refs.~\cite{XuePRD,Xue2023b,Xue2024} ($r=0.022$, $g_Y^2=10^{-12}$, $\Delta x=0.5$).

\section{Kinetic equations and the rates \texorpdfstring{$\GaM$}{GammaM} and \texorpdfstring{$\GaC$}{GammaC}}\label{app:kinetic}

This appendix derives the rates $\GaM$ and $\Gamma_C$ and the charge equation from the couplings to the spacetime vacuum. It follows Ref.~\cite{Xue2026b} [Sect.~4.2, Eqs.~(4.3)--(4.5)].

\paragraph{Couplings and spectral function.} The fluctuations $\delta\Ocal_S$ of the spacetime vacuum, with dynamics $\mathcal L_S$, are characterised by their spectral function
$S_{\mathcal O}(\omega)=\int\dd t\,e^{i\omega t}\langle\delta\Ocal_S(t)\delta\Ocal_S(0)\rangle$ [Ref.~\cite{Xue2026b}, Sect.~4.2 and Appendix~A]. The pairs couple to them through $-\delta\Ocal_S|\Phi|^2$ [Eq.~\eqref{eq:Leff}] and,
in the effective Lagrangian, through $-\tfrac{g_c}{2}\delta\Ocal_S(\Phi^2+\Phi^{*2})$ [Eq.~\eqref{eq:DeltaL}]. With the mode expansion of Sect.~\ref{sec:Leff} and $u_k\simeq e^{-i\int\Omega_k}/\sqrt{2\Omega_k}$, the matrix
element for creating two quanta of momenta $\pm\kk$ is $1/2\Omega_k$ in both cases, times $g_c$ for the same-sign channel. The pair absorbs the energy $2\Omega_k$ from the spacetime vacuum.

\paragraph{Golden rule.} To lowest order the transition rates are
\begin{equation}
  \Gamma_{M,k}^\uparrow=\frac{S_{\mathcal O}(-2\Omega_k)}{4\Omega_k^2},\quad \Gamma_{M,k}^\downarrow=\frac{S_{\mathcal O}(+2\Omega_k)}{4\Omega_k^2};\qquad
  \Gamma_{C,k}^{\uparrow,\downarrow}=g_c^2\,\Gamma_{M,k}^{\uparrow,\downarrow}.
  \label{eq:app_golden}
\end{equation}
If the spacetime vacuum is in a stationary state, the KMS condition $S_{\mathcal O}(+\omega)/S_{\mathcal O}(-\omega)=e^{\omega/T_S}$ defines its temperature $T_S$, and $\Gamma_{M,k}^\downarrow/\Gamma_{M,k}^\uparrow=e^{2\Omega_k/T_S}$.

\paragraph{Pairwise channel and $\GaM$.} With Bose enhancement, the occupation of a mode obeys [Ref.~\cite{Xue2026b}, Eq.~(4.4)]
\begin{equation}
  \dot f_k=\Gamma_{M,k}^\uparrow(1+f_\Phi)(1+f_{\Phi^*})-\Gamma_{M,k}^\downarrow f_\Phi f_{\Phi^*},
  \label{eq:app_kin1}
\end{equation}
identically for $f_\Phi$ and $f_{\Phi^*}$, which conserves the charge. Linearising about the equilibrium occupation $f_k^*$, with $f_\Phi=f_{\Phi^*}=f_k^*+\delta f_k$ and detailed balance $\Gamma_{M,k}^\uparrow(1+f_k^*)^2=\Gamma_{M,k}^\downarrow f_k^{*2}$, a deviation relaxes as
$\delta\dot f_k=-\Gamma_{M,k}\,\delta f_k$, with the per-mode rate $\Gamma_{M,k}=2\big[\Gamma_{M,k}^\downarrow f_k^*-\Gamma_{M,k}^\uparrow(1+f_k^*)\big]=2\Gamma_{M,k}^\downarrow f_k^*/(1+f_k^*)$. Integrating with the weight $\Omega_k$ over the modes and adding the decays gives the cosmic rate equation \eqref{eq:rate} [Ref.~\cite{Xue2026b}, Sect.~4.4, Eqs.~(4.9)--(4.10)], with
the relaxation rate
\begin{equation}
  \GaM=\big\langle\Gamma_{M,k}\big\rangle_{\Omega}=\Big\langle2\,\Gamma_{M,k}^\downarrow\times f_k^*\Big\rangle_{\Omega}\big(1+\mathcal O(f^*)\big)\simeq\frac{\chi m\epsilon}{4\pi},
  \label{eq:app_GaM}
\end{equation}
where $\langle\cdot\rangle_\Omega$ is the average over the modes weighted by $\Omega_kf_k$. The kinetics does not fix the coefficient; the estimate on the right, with its derivation and discussion, is given in
Ref.~\cite{Xue2026b} [Sect.~5.4, Eq.~(5.4)]. The gain term $\GaM\rhoMH$ is the continuous production of neutral pairs, Eq.~\eqref{eq:nprod}.

\paragraph{Same-sign channel and $\Gamma_C$.} The time-derivative coupling $-\dot f\,J^0$ shifts the energies of particles and antiparticles to $\Omega_k\mp\mu$. Creating a same-sign pair $\Phi\Phi$ therefore costs
$2(\Omega_k-\mu)$, and $\Phi^*\Phi^*$ costs $2(\Omega_k+\mu)$. Detailed balance gives the factors $e^{\pm2\mu/T_S}$ of Eq.~\eqref{eq:kin_DQ2}. The same-sign terms of Eq.~\eqref{eq:kin_DQ2} vanish for
$f_{\Phi,\Phi^*}=[e^{(\Omega_k\mp\mu)/T_S}-1]^{-1}$, which gives Eq.~\eqref{eq:deltaeq}. Write $f_{\Phi,\Phi^*}=f^*\pm\tfrac12\Delta$, with $f^*$ the symmetric occupation and $\Delta$ small, and expand to first order in
$\Delta$ and $\mu/T_S$. The pairwise terms cancel in $\dot\Delta$, and
\begin{equation}
  \dot\Delta=-4\Gamma_{C,k}^\downarrow f_k^*\Big[\Delta-\frac{\mu}{T_S}\,2f^*(1+f^*)\Big]\Big(1+\mathcal O(f^*)\Big),
  \label{eq:app_lin}
\end{equation}
using $\Gamma_C^\uparrow(1+f^*)^2=\Gamma_C^\downarrow f^{*2}$ at equilibrium. The charge of mode $k$ therefore relaxes towards $\delta_{\rm eq}=\mu/T_S$ at the rate $\Gamma_{C,k}=4\Gamma_{C,k}^\downarrow\times f_k^*$, the same-sign analogue of $\Gamma_{M,k}=2\Gamma_{M,k}^\downarrow\times f_k^*$ (two quanta per pair). Its mode average is $\Gamma_C$. Since $\Gamma_{C,k}^\downarrow=g_c^2\Gamma_{M,k}^\downarrow$ and
$\GaM$ is set by the same spectral function, $\Gamma_C=\kappa_C\GaM$, with $\kappa_C\propto g_c^2$.

\paragraph{Macroscopic charge equation.} Integrating $\Delta$ over the modes gives the charge density $n_Q=n_\Phi-n_{\Phi^*}$:
\begin{equation}
  \dot n_Q+3Hn_Q=-\Gamma_C\big(n_Q-\delta_{\rm eq}n_M\big).
  \label{eq:app_nQ}
\end{equation}
The pair number obeys $\dd(a^3n_M)/\dd t=(3-2\epsilon)Ha^3n_M$ while it tracks $\rhoMH\propto H^2$. Dividing, $\delta=n_Q/n_M$ satisfies Eq.~\eqref{eq:delta_eq_ode}. In e-folds, with Eq.~\eqref{eq:G2kappa},
this is Eq.~\eqref{eq:charge_grav}.

\section{Numerical methods and parameter values}\label{app:numerics}

\paragraph{Background.} The closed sets of Eqs.~(\ref{eq:closedset}) and [Ref.~\cite{Xue2026b}, Eq.~(8.1)] are integrated in $x=\ln a$ with implicit Runge--Kutta (Radau) or LSODA methods.
\begin{itemize}[leftmargin=1.5em,itemsep=1pt]
\item The relative tolerance is $10^{-8}$--$10^{-10}$, the maximal step $10^{-2}$ e-fold, and events locate $\rho_\Lambda=0$.
\item In Sect.~\ref{sec:transition} the inflationary parameters and the reheating history are iterated until the horizon-matching value of $N_*$ changes by less than
$10^{-4}$.
\item Horizon matching uses $\ln(k_*/a_0H_0)=5.39$ for $k_*=0.05\,{\rm Mpc}^{-1}$, $T_0=2.725$\,K, $g_{\rm eff}=106.75$ and $g_{*0}=3.91$.
\end{itemize}

\paragraph{Parameters.} The main parameter values used in the article are:
\begin{center}\small
\begin{tabular}{@{}lll@{}}
\toprule
quantity & value & source\\
\midrule
$n_s$, $\epsilon$ & $0.965$, $0.0175$ & Planck\\
$H_*$ & $3.15\times10^{-5}(r/0.1)^{1/2}\mpl$ & $A_s$, $r$~\cite{Xue2023}\\
$\chi$, $m_*$ & $9.9\times10^{-4}$, $4.21\,\mpl$ ($r=0.022$) & end of inflation $H=\GaM$\\
$m_{\mathcal M}$, $\epsilon_{\mathcal M}$ & $37.9\,\mpl$, $1.42$ & $\mathcal M$-episode\\
$g_Y^2$ & $10^{-12}$ ($10^{-14}$--$10^{-9}$) & reheating\\
$T_{\rm RH}$ & $6.2\times10^{12}$\,GeV & Sect.~\ref{sec:transition}\\
$N_*$ & $53.5$ & horizon matching\\
$n_B/s$ & $8.7\times10^{-11}$ & CMB, BBN\\
$\delta_{\rm eq}=\mu/T_S$ & $1.4\times10^{-3}$ ($c_B=1$), $4.0\times10^{-3}$ ($c_B=28/79$) & Sect.~\ref{sec:asymM}\\
\bottomrule
\end{tabular}
\end{center}

\bibliographystyle{JHEP}
\bibliography{refs}
\end{document}